\documentclass[%
 aip,
 amsmath,amssymb,
 reprint,%
]{revtex4-1} 

\PassOptionsToPackage{hyphens}{url}\usepackage{hyperref}

\usepackage{graphicx}% Include figure files
\usepackage{dcolumn}% Align Tab. columns on decimal point
\usepackage{bm}% bold math
\usepackage[utf8]{inputenc}
\usepackage[T1]{fontenc}
\usepackage{mathptmx}
\usepackage{xcolor}

\urldef{\footurl}\url{https://github.com/strangeli/controlled-multi-timescale-powergrid}

\DeclareMathOperator{\sgn}{sgn}

\newcommand{\specialcell}[2][l]{\begin{tabular}[t]{@{}l@{}}#2\end{tabular}}

\begin{document}

 %\preprint{AIP/123-QED}

\title[]{A multiplex, multi-timescale model approach for economic and frequency control in power grids.}
% Force line breaks with \\

\author{Lia Strenge}
\affiliation{ 
Control Systems Group at Technische Universit\"at Berlin, Germany
}%
\author{Paul Schultz}
\author{Jürgen Kurths}
\affiliation{%
Potsdam Institute for Climate Impact Research, Germany
}%
\author{Jörg Raisch}
\affiliation{ 
Control Systems Group at Technische Universit\"at Berlin, Germany
}%
\author{Frank Hellmann}
 \email{hellmann@pik-potsdam.de}
\affiliation{%
Potsdam Institute for Climate Impact Research, Germany
}%

\date{\today}% It is always \today, today,

%%%%%%%%%%%%%%%%%%%%%%%%%%%%%%%%%%%%%%%%%%%%%%%%%%%%%%%%%%%%%%%%%%%%%%%%%%%%%%%%
\begin{abstract}
Power systems are subject to fundamental changes due to the increasing infeed of decentralised renewable energy sources and storage. The decentralised nature of the new actors in the system requires new concepts for structuring the power grid, and achieving a wide range of control tasks ranging from seconds to days. Here we introduce a multiplex dynamical network model covering all control timescales. Crucially, we combine a decentralised, self-organised low-level control and a smart grid layer of devices that can aggregate information from remote sources. The safety-critical task of frequency control is performed by the former, the economic objective of demand matching dispatch by the latter.
Having both aspects present in the same model allows us to study the interaction between the layers. Remarkably, we find that adding communication in the form of aggregation does not improve the performance in the cases considered. Instead, the self-organised state of the system already contains the information required to learn the demand structure in the entire grid.
The model introduced here is highly flexible, and can accommodate a wide range of scenarios relevant to future power grids. We expect that it is especially useful in the context of low-energy microgrids with distributed generation.
\end{abstract}

\maketitle

\begin{quotation}
Highly decentralised power grids, possibly in the context of prosumer systems, require new concepts for their stable operation. We expect that both self-organised systems as well as intelligent devices with communication capability that can aggregate information from remote sources will play a central role. Here we introduce a multiplex network model that combines both aspects, and use it in a basic scenario and uncover surprising interactions between the layers.
\end{quotation}

\section{Introduction}
Power systems are subject to fundamental changes caused by the increasing infeed of decentralised and fluctuating renewable energy sources. 
One key change is that the conventional energy producers, i.e., big central power plants, are currently also the locus of control resources for the power grid. 
The challenge facing future power grids lies in achieving a stable and robust operation of the grid without such centrally-controlled actors.

In general, the objective for a stable operation is to maintain the frequency and voltage of the system and to keep the system in an economically desired state. To do so, it is necessary to achieve an instantaneous balance between electricity generation and consumption. Any imbalance is directly linked to a deviation of the grid frequency from its nominal value (50 Hz/ 60 Hz). The control and stabilisation of frequency is traditionally divided into primary, secondary and tertiary control. 
They respectively address instantaneous frequency stabilisation (primary), i.e., keeping frequency within given bounds, restoring the nominal frequency (secondary), and achieving or restoring a desired economic state (tertiary). 
These three control layers also typically come with a temporal hierarchy, with primary control acting at the scale of seconds, secondary in minutes and tertiary in quarter-hours.

In addition, the corresponding tasks require an increasing level of explicit communication and coordination of the actors, who participate in various markets to ensure sufficient control resources in an economically feasible way. 
While primary control might be achieved through automated reactions to frequency deviations, controls on slower timescales are typically subject to active human decisions. 

The power required to operate the safety-critical primary and secondary control as well as to balance out unforeseen load variations, is typically held in reserve, and has to be provided at short notice. Both reserved capacity and energy drawn cause costs. On the other hand, the energy required to service the expected load is bought days, weeks or even years in advance, based on past experience. 
This energy can be dispatched by the cheapest provider, and technical constraints play less of a role here. It can thus be expected to be cheaper overall than primary or secondary control energy. 
Following a fault, it is the role of tertiary control to restore this economically favourable state.

Control concepts for (prosumer-based) microgrids, which could form a key part of future grid designs, especially in emerging markets, follow the same hierarchy of tasks \cite{Guerrero2010,schiffer2015modeling}. 
Again, primary and secondary control are required for the proper functioning of the grid itself, whereas tertiary control (also called energy management) chooses the economically desirable source of energy.
With this context in mind, we explore scenarios for achieving frequency stability as well as 
economically optimal balancing.
We focus on distributed, self-organised control actors, assuming that an analogous detailed market design for single microgrids is not feasible due to their small sizes, decentralised power provision and low inertia. 

In future power grids, the control tasks will have to be performed by new distributed actors. 
A key question in their grid design is how much coordination these actors will require and how much can be achieved through self-organised means. 
To obtain a system that can function in the face of communication failures, it is natural to require that primary and secondary control should be achieved in a fully decentralised and self-organised fashion. 
On the other hand, tertiary control is an optimisation task that can make use of communication and coordination infrastructure safely.

To understand whether such a communication and coordination layer is required, and how it performs with respect to control and stabilisation, it is necessary to study the interaction of the different layers of the control hierarchy, which are typically studied only separately.
 %Specifically for tertiary control there are a variety of approaches. %\cite{simoes2006intelligent} uses neural networks for forecasting and optimization for distributed generation.. 
Most literature on hierarchical control is reviewing existing approaches on their respective timescales without explicitly studying their interaction \cite{bidram2012hierarchical,dorfler2014plug, olivares2014trends,Aamir2016, xin2015decentralized,han2016review,li2017fully}.

In this paper we introduce a model for the hierarchical operation of a power grid that aims to achieve two goals,
robust control in the face of communication failures, and some notion of an economically optimal dispatch under operational constraints.
We consider the power grid as a two layer network \cite{Gao2011a,Gale2014,Kivela2014}, where the layers are given by (i) the physical electricity network together with a fully decentralised real-time distributed primary/secondary control of the grid frequency, and (ii) an energy management layer. 
If we allow for communication in the energy management layer, this can be described as a multiplex network \cite{DeDomenico2013a,Nicosia2013,Boccaletti2014} of a physical and a control layer, i.e., both layers have an identical set of nodes. 
In the latter, links correspond to a directed information transfer between controllers. 
This model allows us to study the interaction of timescales ranging from seconds to days.

We use this model to analyse a simple but illustrative scenario, where a subset of nodes in the system has the ability to dispatch energy in hourly intervals, and optimise for an unknown periodic background demand that is inferred from the control actions required by the primary/secondary layer. 
Using this scenario we can compare the effect of various communication strategies.

In order to study the performance of various approaches, we make use of probabilistic methods \cite{menck2013basin, menck2014dead, schultz2014detours, hellmann2016survivability, schultz2018bounding, lindner2019stochastic, kim2016building, kim2018multistability, wolff2018power}, that is, we define a scenario ensemble and evaluate the expected performance of the system with respect to the ensemble average by sampling over the ensemble. 
This approach allows us to study the properties of the system for the whole ensemble, rather than in individual case studies. 
As our main aim here is to introduce the model and understand qualitatively the performance of the control layers, our setup is rather conceptual and does not capture real power systems in detail. 
While the inspiration is drawn from the context of prosumer-based microgrids, we consider this to provide a broader perspective as well.

\subsection{Multiplex aspects of power grids} 

The study of multilayer networks as (dynamical) systems with an additional mesoscale structure experienced an active development in recent years (see \citet{Kivela2014} for a review). 
Subsequently, a variety of statistical network characteristics (e.g., \citet{Donges2011}) has been developed to quantify the multilayer structure.
A special multilayer structure is the multiplex (or multilevel \cite{Criado2012}) network which we employ in our model. Nodes are identical across the layers, hence, the topology between the layers is fixed. 

We identify the network layers with their different functional roles within the system, i.e.,  electricity distribution and control. 
Likewise, the coupling mechanism is different and given by the physical power flow and communication, respectively. 
Introducing an interdependence between the layers affects the overall system's resilience to failures, a popular example is the 2003 Italian blackout\cite{Schneider2011} with a multilayer cascading failure. 
In particular, it has been shown that the interconnection of different networks can promote network breakdown in discontinuous first-order transitions \cite{Buldyrev2010b,Vespignani2010}. 

% Critically, a multiplex network's resilience is not a monotonous function of the connectivity between layers but might be optimal at an intermediate level \cite{Brummitt2012}.

Note that there are a number of works that study consensus-based methods for achieving certain objectives (see, e.g., \cite{krishna2019consensus,mahmoud2015review}) by introducing an additional communication layer. 
These multiplex networks differ in various ways from the setup studied here. Most importantly in the fact that the layers cooperate on a single control objective and it is mostly secondary control \cite{wu2014secondary,han2016review} or quasi-stationary tertiary control \cite{Mao2019} that is considered.

%Therefore, adapted methods for modeling and simulation of power grids with respect to structuring and control are required. 
%The concept of prosumer-based microgrids gives the opportunity to rethink structuring and operation of power systems from 
%scratch \cite{schiffer2015modeling}.  In analogy to traditional power grids, the hierarchical control in microgrids is typically 
%also divided in primary, secondary and tertiary control, also called energy management \cite{Guerrero2010}. The operation is usually hierarchically structured along timescales and markets which do not interact. In microgrids, the layer analogy refers to the control tasks, hence, primary control is instantenous frequency stabilization whose steady state error is compensated by secondary control. Tertiary control accounts for all tasks related to economic dispatch. In our case, a decentralisedized leaky integral controller for combined primary and secondary control achieving banded frequency restoration is used on the lower layer\cite{doerfler2018leakyintegrator}. 
%\texttt{(Lia: Like this?)}

\subsection{Energy management}

For a decentralised primary and secondary control we make the most simple choice, and use a lag element, which can also be  described as adapted distributed proportional-integral (PI) control. 
The main adaptation, discussed thoroughly in \cite{doerfler2018leakyintegrator} is that the integral controller includes an exponential decay term. 
While this means that there remains a residual steady state error, it makes the setup robust to unavoidable systematic errors in the implementation of the integrator.

The question of how to model energy management is far more challenging, and less settled. 
A variety of approaches\cite{weidlich2008critical, ringler2016agent, mureddu2018extreme} model energy markets directly. 
For tertiary control (here referring to the technical implementation) directly, we are not aware of any general models, though more concrete studies for the microgrid context exist\cite{claessen2014comparative}. 
There are also various works studying the relationship to congestion management and frequency control\cite{zhao2016unified}. 
Our approach here is to side step the question of how exactly the dispatch is chosen, but instead focus on studying the steady state emerging once the economic optimisation (given the available information) has been performed. 
We do so by introducing an iterative learning control (ILC)\cite{amann1998predictive, bristow2006survey} that considers the previous days performance and attempts to iteratively improve it by scheduling a different dispatch for the next day. 
As the consumption/production fluctuates from day to day, this iterative process should converge to an optimum dispatch. Note that the energy scheduled for the next day is known a day ahead but not generated a day ahead.

ILC is a control method which can be applied to track a periodic output or reject periodic disturbances. The error is reduced over the iteration cycles and it can easily be combined with feedback controllers. ILC has previously been used in power systems in other contexts, mainly for inverter control, e.g., \cite{zeng2013topologies, teng2014repetitive}. %\texttt{(I have this in more detail if needed)}.
In addition, ILC is applied to an uninterruptible power supply\cite{Aamir2016}  and for optimal residential load scheduling\cite{chai2016}. In building automation, data-driven methods for demand response in the residential building sector are taken into account\cite{bampoulas2019self}; ILC also addresses frequency control with high penetration of wind integration\cite{Guo2016}; it is further applied to energy management in electric vehicles \cite{Guo2019}. 
Hence, most of the literature combining energy management and ILC focus on single nodes in a grid without emphasis on the overall grid perspective.
However, a review on ILC for energy management in multi-agent systems states that the applicability of ILC to the topic including physical constraints has a high research potential due to its (periodic) disturbance rejection capacity and distributed architecture for large-scale systems\cite{nguyen2016iterative}.  
ILC for physically interconnected linear large-scale systems is studied and applied to economic dispatch in power systems\cite{xu2013iterative}. 

In larger grids we expect that this learning would be replaced, for example, by a market-based system. 
However, in autonomous microgrids, without the resources necessary to implement a market-based solution, the ILC itself is a viable way of choosing dispatch. 
We leave a detailed discussion of the design, as well as a proof of linearised asymptotic stability in the iteration domain of the ILC in such a scenario to a companion paper \cite{Strenge2019}.

This paper is structured as follows. In Section \ref{s:modeling}, the overall model with two control layers is presented. In Section \ref{s:performance}, we compare the performance of the system for different multiplex topologies with sampling based numerical experiments. Finally, in Section \ref{s:discuss}, we discuss our main result that an additional communication layer is not needed in the proposed setting and suggest further research directions. 

\subsection*{Notation} 
Let $\mathcal N$ be the set of nodes in the electricity network. 
Then we have the two graph layers. Firstly, the electricity network 
$\mathcal{G}=(\mathcal{N}, \mathcal E)$ is an undirected graph, i.e., $\mathcal E\subseteq\mathcal N\times \mathcal N$ with  $(i,j)\in \mathcal E \Leftrightarrow (j,i) \in \mathcal E$. 
Secondly, the communication layer is represented by a directed graph $\mathcal{G}^{C}=(\mathcal{N}, \mathcal E^{C})$, with a bipartition of $\mathcal{N}$ into $\mathcal{N}^{C}$ with higher-layer control present
and $\mathcal{N}^n$ without higher-layer control. $\mathcal{N}^{C} \dot \cup \mathcal{N}^{n} =\mathcal N$ 
where $\dot \cup$ is the disjoint union.
%i.e., $i\in\mathcal N^C\Rightarrow j\notin\mathcal N^{C}, (i,j)\in \mathcal E \vee (j,i) \in \mathcal E$. 
Note that $\mathcal E^{C} = \{(i,j)| i \in \mathcal N, j\in  \mathcal N^C\}$ (edges directed from $i$ to $j$) is not necessarily a subset of $\mathcal E$ and data is available from all nodes in $\mathcal N$. We call $\mathcal S\subset \mathcal N$  a maximal independent set in $\mathcal N$, i.e.,$\forall i \in \mathcal N$: $i\in \mathcal S \vee N(i)\cap\mathcal S \neq \emptyset$ where $N(i)$ denotes the neighbors of $i$. We label the nodes $j\in\mathcal N=\{1,...,N\}$. card($\cdot$) denotes the cardinality of a set and $\times$ the Cartesian product of two sets.

\section{Modelling}
\label{s:modeling}

As noted above we use a straightforward and well studied model for providing the basic frequency control of the system with bounded frequency deviation, cp. \cite{schiffer2015modeling, doerfler2018leakyintegrator}.

%\subsection{System model}
\subsection{Lower layer}

\begin{figure}
\includegraphics[width=\columnwidth]{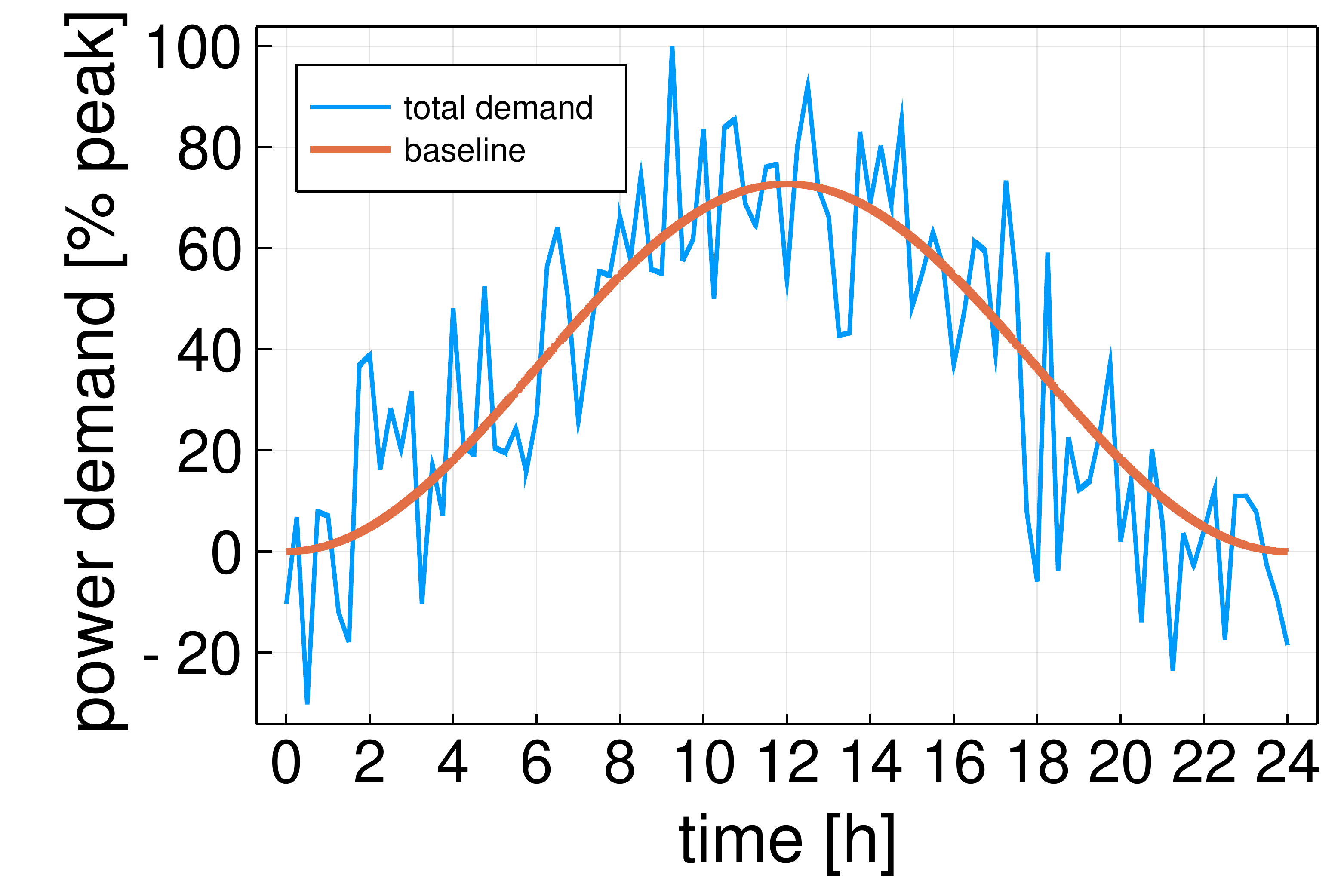} 
\caption{Exemplary demand curve (periodic and fluctuating component)}
\label{fig:demand}
\end{figure}

Our aim is to use a conceptual model for the dynamics of a single node that can capture a variety of behaviours. In keeping with the inspiration of a fully distributed microgrid, we assume that all nodes have control capability. We further assume that the distributed control mimics the relationship between synchronous frequency and the power balance of generation and demand found in traditional synchronous machines \cite{Hill2006,Machowski2011}. Neglecting voltage dynamics this leads us to the formulation of the Kuramoto model with inertia\cite{schiffer-cond,schiffer2015modeling}, with the input power (i.e. basically the natural frequency of the oscillating units) controlled by the distributed control. Further we assume that there are some nodes that allow for slower, dispatchable energy that is controlled by the ILC in the higher layer.

The open-loop system equations for the nodes $j$ are then given by
\begin{subequations} 
\label{eq:mn_plant}
 \begin{align}
 M_j \ddot \phi_j(t) &=  - P^d_j(t) + P^{LI}_{j}(t) + P^{ILC}_{j}(t) + F_j(t), \\
  F_j(t) &= - \sum_{k\in\mathcal N} V_jV_k Y_{jk}\sin\left(\phi_j(t) -\phi_k(t) \right),
 \end{align}
\end{subequations} 
with the time $t\in \mathbb R$; $\phi_j$ is the voltage phase angle  of node $j$ in the co-rotating frame and $ \omega_j := \dot \phi_j$ its instantaneous frequency deviation from the rated grid frequency.
$ F_j $ denotes the AC power flow  from neighbouring nodes under the assumption of purely inductive lines. In general, our approach is not limited to this assumption though. The parameters are the effective inertia $M_j$; the steady-state voltages $V_j$ and the nodal admittances $Y_{jk}$, which encode the network topology. 

The system is driven by the balance between power demand and generation, here, $P^{ILC}_{j}$ and $P^{LI}_{j}$ are the power dispatched by the higher layer (ILC) if available, and the distributed control (LI),
respectively. While we assume that $P^{LI}_{j}$ can be set arbitrarily, $P_j^{ILC}$ has restrictions. That is, $P_j^{ILC}$ has to be chosen to lie within the achievable behaviour $P_j^{ILC} \in \mathcal{B}^{ILC}_j$ of the dispatchable energy at the node $j$. This can encode a variety of constraints such as minimal run times, maximum ramp rates, and finite storage. In order to mimic the behaviour of trading markets, with their hourly or quarter-hourly dispatch, we  here take $\mathcal{B}^{ILC}_j$ to be the space of functions that are constant during each hour.

The energy demand $P^d_j$ is a priori unknown. In order to efficiently study the hierarchical control performance with regard 
to communication structure and the stochastic nature of the system, we use well-defined 
synthetic demand curves. In particular $P^d_j = P^p_j + P^f_j$ is composed of $P^p_j$, which is a periodic baseline demand with randomly 
selected amplitude at each node (period $T_d$ of a single day), and $P^f_j$ is composed of 
additive white noise with zero mean with a piece-wise linear interpolation in intervals of 15 minutes. This is visualised in Fig.~\ref{fig:demand}. See Appendix~\ref{a:demand} for an explicit formulation. 
 While for simplicity we here use a traditional separation of dispatch and demand, the model naturally accommodates fluctuating production as well. Note that all power-related quantities here are scaled with a rated power.

The decentralised control in the lower layer is responsible for primary and secondary control tasks which is 
frequency stability and restoration.
Hence, the addressed control objective for nodes $j\in\mathcal N$ 
is to achieve a bounded frequency deviation:
\begin{equation}
\forall_t\forall_{j\in\mathcal N} \, \omega_j(t) \in [\omega_{min},\omega_{max}]  \label{eq:boundedf} \;,
\end{equation}
where we consider $\omega_{min}=49.8$ Hz $= 306.62$ rad/s and $\omega_{max}=50.2$ Hz $= 315.42$ rad/s.
The lower-layer controller \cite{doerfler2018leakyintegrator} is chosen for this purpose. It has a small steady state error but it avoids instability caused by parallel integrators.  The dynamics are as follows: 
\begin{eqnarray}
\label{eq:LI}
  P^{LI}_{j}(t) &=& -k_{p,j} \omega_j(t) + \chi_j(t),\\
  T_j \dot\chi_j(t) &=& - \omega_j(t) - k_{I,j} \chi_j(t)	, 
\end{eqnarray}
where  $\chi_j$ is the controller state, $T_j$, $k_{I,j}$ and $k_{P,j}$ are parameters of the lower layer controller.

The choice of the control parameters should be in compliance with known design criteria \cite[Corollary 4]{doerfler2018leakyintegrator}. In order to specifically choose the parameters within the remaining degrees of freedom, we select them by probabilistic methods. The results are shown in Figs. \ref{fig:cp_kp1} to  \ref{fig:cp_kp3}  in Appendix \ref{a:control_parameters}.
Hence, we choose $k_{P,j} = 525$ s/rad and $k_{I,j} = 0.005$ rad/s. 
The steady state error \cite[Corollary 4]{doerfler2018leakyintegrator} for the maximum possible demand in this setting is 0.001655 rad/s.

Concretely, we use two complementary measures of the performance of the primary and secondary control, the maximum observed frequency deviation 

\begin{equation}
\omega_{top} := \max_{j\in\mathcal N} \, \max_{t\in\left[t_{obs,start}; t_{end}\right]}\, \omega_j(t)
\end{equation}
and the exceedance, which is the total time that the frequency is out of bounds within an observed interval, i.e., 
%\begin{equation}
%\sum_{k=1}^{n_{\Delta t}} (\Delta t)_k \text{ with }\omega_j(?) \notin [\omega_{min},\omega_{max}] \forall j\in\mathcal N, 
%\end{equation}
\begin{equation}
exc_j := \frac{1}{t_{end}-t_{obs,start}} \int_{t_{obs,start}}^{t_{end}} \Theta(|\omega_j(t)| - \Delta\omega) dt.
\end{equation}
In this terminology, $exc>0$ would mean that objective Eq.~\eqref{eq:boundedf} has not been achieved. However, to better resolve performance differences between the different control designs discussed below, 
we choose a tighter bound $\Delta \omega = 0.0005$ rad/s for the defining critical frequency deviation.

Since the learning control introduced in the following section takes some time to converge to a steady state, 
we discard the transients and focus on the performance over an interval $\left[t_{obs,start}; t_{end}\right]$.

%\subsection{Supervisor}
\subsection{The ILC control layer}
%The high-level control for the specification 
%\begin{equation}
%$ E^n_{ctrl,j} = \int_{t_n}^{t_{n+1}} P^{LI}_{j}(\tau) d\tau \leq E_{max,j}$
%\end{equation} 
%is based on the aggregation 
% \begin{align}
% E_{\pm,j}^n &= \frac{1}{2} \left(\tilde{E}^n_j \pm E^n_j \right) ,\quad E_{\pm,j}^n > 0 \\
% E^n_j &= \int^{t_{n+1}}_{t_n}u_j(u^n, t; P^n, D^n_s,D_f)~dt \\
% \tilde{E}^n_j &= \int^{t_{n+1}}_{t_n} \vert u_j(u^n, t; P^n, D^n_s,D_f) \vert~dt
% \end{align}

The aim of the higher-layer controller is to achieve a state that is in some sense economically optimal. As stated above, we study the equilibrium state rather than the convergence to that state. We consider the former a sensible stand-in for other methods that try to achieve an economically optimal situation. The design and performance of the ILC itself as a method for microgrids is treated in a companion paper\cite{Strenge2019}.

Concretely, the higher-layer controller looks at the previous day and adjust the dispatch chosen from $\mathcal B^{ILC}_j$ in such a way as to minimise the overall system cost. 
Therefore, the economic objective translates to
minimising 

\begin{equation}
 \mathcal{C}_{total} = 
 \sum_{j=1}^N \int_{t_{obs,start}}^{t_{end}}  
 \lambda \vert P^{LI}_j(\tau) \vert +  
 (1-\lambda) \vert P^{ILC}_j(\tau)\vert d\tau \; ,\label{eq:cost}
\end{equation}
 with the (here constant) cost factor $\lambda \in (0,1]$. Hence, $\tfrac{\lambda}{1-\lambda}$ is a notion of the price relation between $P^{LI}$ and $P^{ILC}$.
As noted above, the modelling assumption here is that power planned a day in advance is technically less challenging, and thus cheaper than control power that needs to be provided as an instantaneous reaction, and thus $\lambda > 0.5$. This is aligned with current market pricing where day-ahead markets trade energy at a much lower price than primary and secondary control energy markets. A similar relation can be expected in microgrids. Fig. \ref{fig:lambda} in Appendix \ref{a:update} shows how the average overall system cost changes with the cost factor $\lambda$. For our analysis, we choose $\lambda = 0.8$.

An update is chosen within $\mathcal{B}^{ILC}_j$, and the next day's $P^{ILC}_j$ are adjusted accordingly, possibly aggregating the updates from communicating nodes as well. Concretely we adjust $P_j^{ILC}$ proportionally to
\begin{equation} 
E^h_{ctrl,j} = \lambda \int_{t_h}^{t_{h+1}} \text{sgn}(P^{LI}_j(\tau)) d\tau + \bar T(\lambda -1),\label{eq:update_int}
\end{equation}
%$$E^h_{ctrl,j} = \int_{t_h}^{t_{h+1}} P^{LI}_j(\tau) d\tau,$$
where $t_h=(h-1)t$ with $t\in[t_{start},t_{end}]$, $t_{start}, t_{end}\in\mathbb R_{\geq0}$, is the beginning of hour $h\in\mathbb N$ and sgn($\cdot$) the sign function and $\bar T = 3600$.  The iteration step Eq.~\eqref{eq:update_int} we perform is motivated in Appendix \ref{a:update} and minimizes the cost Eq.~\eqref{eq:cost} under certain assumptions.

Concretely, in the scenario where all nodes have dispatchable energy, and the ILC only updates based on the local information, this implies that we have for each node $j = 1 ,..., N$ and each hour $h$
 \begin{equation}
 P^{ILC, h}_j = P^{ILC, h-24}_j + \kappa_j E^{h-24}_{ctrl, j} \;,
 \end{equation}
where $P^{ILC, h}_j$ is the value of the hourly constant $P^{ILC}_j(t)$ for 
 $t_h \leq t < t_{h+1}$, that is, during the hour starting at $t_h$, and 
$\kappa_j \in \mathbb R$ is the learning gain. 

Let us now consider the genuine multiplex case, where we allow for communication, that is, the ILC nodes aggregate information about the expended control energy at different nodes, and update using the total. Then, using the adjacency matrix $A^{C}$ of the communication layer, we get
 \begin{equation}
 P^{ILC, h}_j = P^{ILC, h-24}_j + \frac{\kappa_j}{d_j + 1} \left(E^{h-24}_{ctrl, j} +  \sum_{k\neq j} A^{C}_{jk} E^{h-24}_{ctrl, k}\right)\;, \label{eq:ILC}
 \end{equation}
for all $j \in \mathcal{N}^{C}$, where $d_j = \sum_{k\in\mathcal N} A^{C}_{jk}$ is the degree.
In terms of control, this means we use a P-type (i.e., proportional) ILC controller. A Q-filter, implying a forgetting filter regarding previous and upcoming hours of a day or a more sophisticated interaction between the nodes, is not considered here, hence $Q=I$ where $I$ is the identity matrix. A Q-filter may become beneficial if the demand amplitudes vary with the days. Furthermore, we simplify and use the same learning gain for all nodes, i.e., $\kappa_j = \kappa$ for all $j\in\mathcal N$.

\section{Performance comparison for different multiplex topologies}
\label{s:performance}

Our main focus is to study whether the communication network is required in order to achieve sensible economic outcomes, or whether the decentralised robust operation of the grid implicitly carries enough information between the nodes to achieve an acceptable. outcome without added communication infrastructure.

\subsection{Grid ensemble}

As noted above, we apply a probabilistic approach. That is, we define a class of grids consisting of a random peak demand per node as well as random topology, and evaluate the expected performance of the design for this grid class. For simplicity, and to focus on the effects of the topology in the higher layer, we use a simple random regular graph with degree three for the topology in the lower layer. This is not intended to be a realistic choice for most power grids, but provides a homogeneous backdrop on which the higher layer can operate. The demand amplitudes are chosen uniformly at random. For details on the demand model, see Appendix \ref{a:demand}.

\begin{table}
\caption{Simulation parameters (node $j = 1,...,N$). \\W/W units refer to scaled quantities.}
\label{tab:parameters}
\vspace{2mm}
\begin{tabular}{|l|l|l|p{3cm}|}\hline
\textbf{Parameter}  & \textbf{Value} & \textbf{Unit} & \textbf{Description}\\
\hline
\hline
$k_{I,j}$  & 0.005 & (W rad)/(Ws) & lower layer control parameter\\
\hline
$k_{P,j}$  & 525 & Ws/(W rad) & lower layer control parameter\\
\hline
$\kappa_j$ & 0.15/3600 &1/s& learning parameter \\ 
\hline
$M_{j}$ & 5 & W s$^2$/(W rad) &  inertia \\
\hline
$N$ & 24 & - & number of nodes \\
\hline
$1/T_{j}$  & 1/0.05 &1/rad& lower layer control parameter \\
\hline
$Y_{jk} V_j V_k$ & 6 & W/W  &  coupling \\
\hline
$\Delta \omega$ & 0.0005 & rad/s  & frequency threshold for the exceedance\\
\hline
$t_{end}$  &  50 & days & number of simulated days\\
\hline
$t_{obs,start}$  &  40 & days & start of observation interval\\
\hline
$S$ & 100 & - & number of simulations in one experiment\\
\hline
\end{tabular}
\end{table}

We consider a sample size of $S=100$ grids \`a $card(\mathcal{N})=24$ nodes with random demand from the ensemble and then integrate the system for 50 days. 
Investigation of individual trajectories reveals that this is highly sufficient to achieve equilibrium for the ILC in all cases considered
(compare Appendix \ref{a:trajectories}; Figs. \ref{fig:timeseries} to \ref{fig:timeseriesIV} which show example trajectories for the different higher-layer scenarios).
 We then use the performance on the last ten days to study the steady state properties. 
  All initial conditions are set to zero, i.e. the ILC update sets in after the first day. 
  Find all relevant simulation parameters in Tab. \ref{tab:parameters}.

We evaluate the expected value of three quantities already introduced above. First, the maximum frequency deviation $\omega_{top}$ to see how far the system deviates from the desired frequency, second, the frequency exceedance $exc_j$ which indicates the quality of the control achieved. The third quantity is the total cost of higher-layer and lower-layer control energy in the system $\mathcal C_{total}$ defined in Eq.~\eqref{eq:cost}. 

\subsection{Higher-layer topologies}

We consider five different topologies chosen to illustrate different designs of communication and control infrastructure.

The first baseline scenario (scenario $\mathbf{0}$) is to study the performance of the decentralised control by itself, without any ILC, that is the higher layer is simply the empty graph, $\mathcal{N}^{C} = \emptyset$.

The second baseline scenario (scenario $\mathbf{I}$) is to assume that every node has the ability to dispatch energy, and optimises to satisfy its demand locally, without taking the neighbours into account, $\mathcal{N}^{C} = \mathcal{N}$ but $\mathcal E^{C} = \emptyset$ and thus $A^{C} = 0$.

\begin{figure}
\centering
\includegraphics[width=\columnwidth]{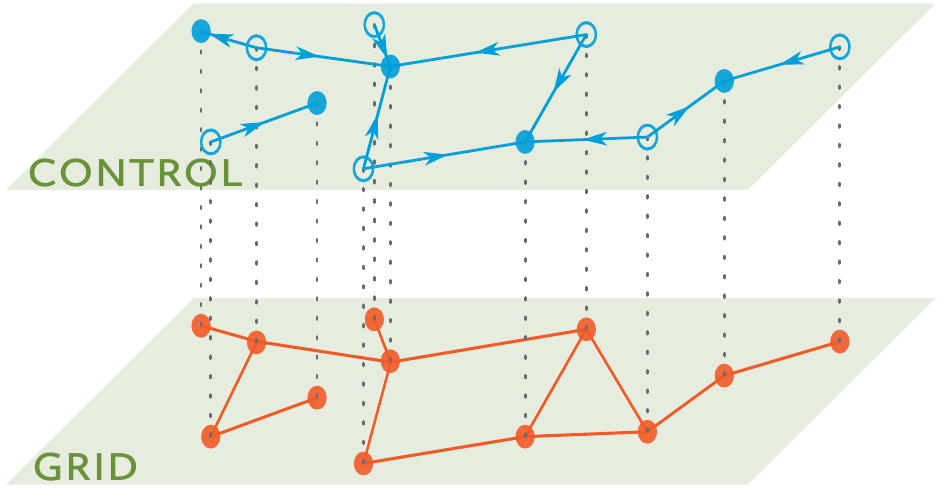}
\caption{An illustration of the multiplex network consisting 
of a physical layer ("grid", bottom) and a control layer ("control", top). Directed edges are indicated with arrows. The vertical dashed lines identify the nodes in the two layers. In the upper layer, filled
circles indicate the nodes $v\in\mathcal{N}^C$ where control is present/active.}
\label{fig:network}
\end{figure}

The three main scenarios we want to consider take ILC at a subset of nodes in such a way as to mimic three different potential designs. The first two (\textbf{II} and \textbf{III}) scenarios assume that the location of the dispatchable power is chosen in some sense to be central in the underlying physical grid. To model this, we construct a maximal independent set $\mathcal N^{C}$, that is, a (non-unique) maximal set of vertices such that 
two nodes in $\mathcal N^{C}$ are never adjacent. Hence, every node in the graph $\mathcal{G}^{C}$ is either in $\mathcal N^{C}$ or neighbour to one node in $\mathcal N^{C}$, cp. Fig. \ref{fig:network} (top). The edges are directed, accounting 
for the directed information
transfer. By assigning the ILC to $\mathcal N^{C}$, every other node in $ \mathcal N^n$ is adjacent to a node with dispatchable energy. With this set of nodes we can now define and compare the scenarios $\mathbf{II}$ and $\mathbf{III}$. Scenario  $\mathbf{II}$ has communication from these neighbors and scenario  $\mathbf{III}$ has no communication. In the former case we have a directed communication graph with the adjacency matrix elements $A^{C}_{jk} = 1$ if $j \in \mathcal N^{C}$ and $Y_{jk} > 0$. Otherwise we set $A^{C}_{jk} = 0$.

This can be seen as a highly conceptual model where local regions in the network are responsible for the energy balance in their respective areas. To contrast this with a random topology, we finally consider the case (scenario $\mathbf{IV}$) that the ILC is positioned at half the nodes at random, and communicate at random with three other nodes.

The three scenarios $\mathbf{II}$, $\mathbf{III}$ and $\mathbf{IV}$ thus represent sparse ILC with no communication, structured communication and random communication. The scenarios are summarised in Tab. \ref{tab:exp}.

%We consider 5 different cases listed in Tab. \ref{tab:exp}. Exemplary networks of the different cases are shown in Fig. \ref{fig:network}. Fig. \ref{fig:timeseries} contains exemplary trajectories for the overall model  Eq.~\eqref{eq:mn_plant}, Eq.~\eqref{eq:LI} and Eq.~\eqref{eq:ILC}. It shows the lower layer control energy for 24 nodes for one simulation for 5 days for the cases 0-IV. It can be observed that the integrated control energy for an empty graph $\mathbf{0}$ rises since there is no update setting it to zero each hour. Local ILC at all nodes $\mathbf{I}$ learns the periodicity after the first day at all nodes. ILC at vertex cover $\mathbf{III}$ learns the periodicity at certain nodes. ILC at vertex cover with neighboring communication (II) and ILC at random nodes with random communication $\mathbf{IV}$ seems to switch the sign for the control energy at certain nodes. We  observe these performance differences in more general forms by the performed numerical experiments.
%
%\texttt{General remark (Lia): Why does the ILC learn within one day all the time? On the poster at the general meeting, we had a leraning effect over several days. By then, we only integrateds $\chi$ and not $P_{LI}$. It would look nicer to see a learning but on the other hand, fast learning is good... In the other paper, we analyze this in more detail.}

\subsection{Results of the numerical experiments}

\begin{figure}[!htp]
\centering
\includegraphics[width=\columnwidth]{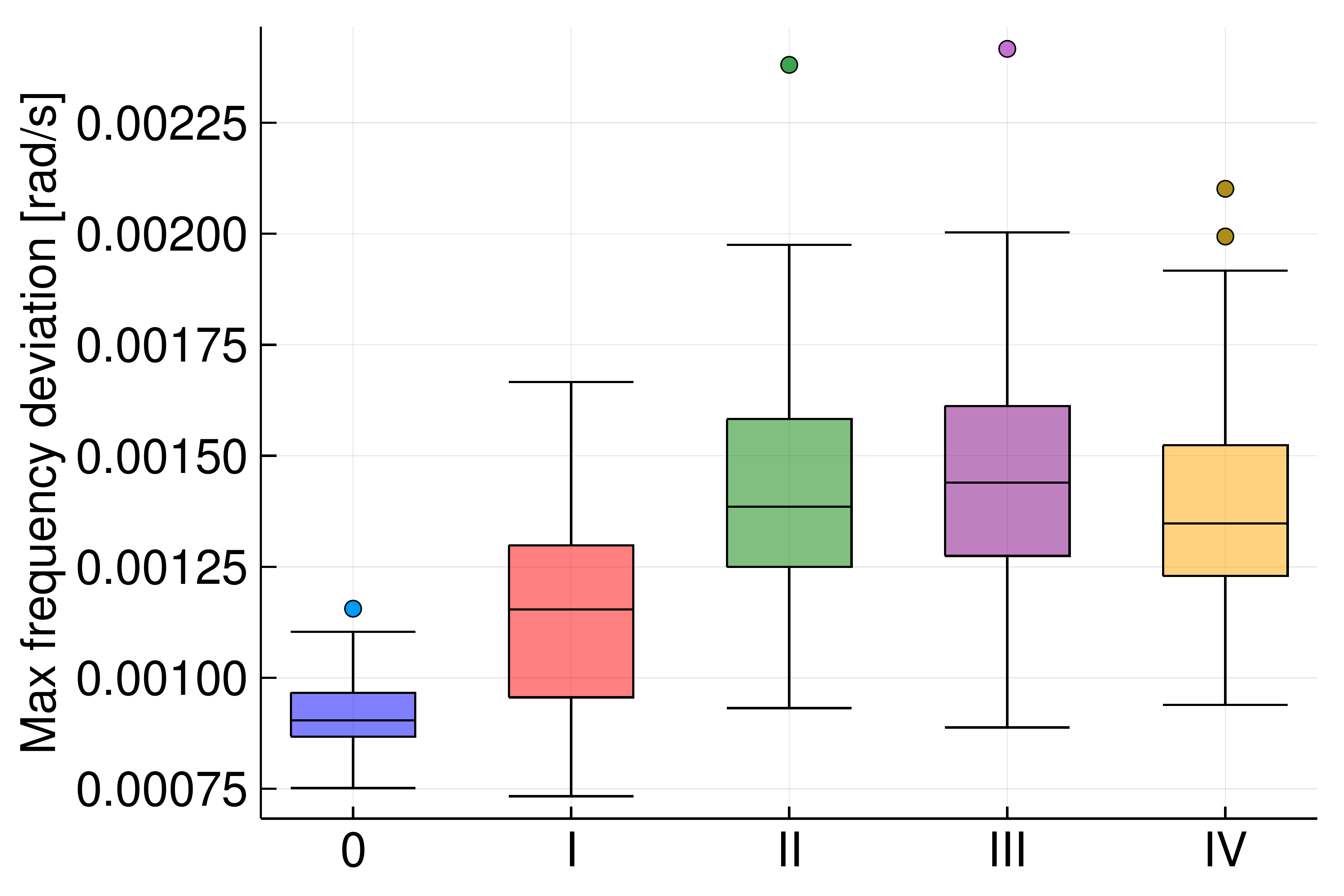}
\caption{The maximum frequency deviations of the nodes for the various higher layers. The box plots show the quartiles and outliers of the system. The coloured box covers the second and third quartile, the middle line gives the median. The T bars give the extrema of the distribution up to outliers. See Tab.~\ref{tab:exp} for the scenario definitions.}
\label{fig:freq_dev}
\end{figure}

\begin{figure}[!htp]
\centering
\includegraphics[width=\columnwidth]{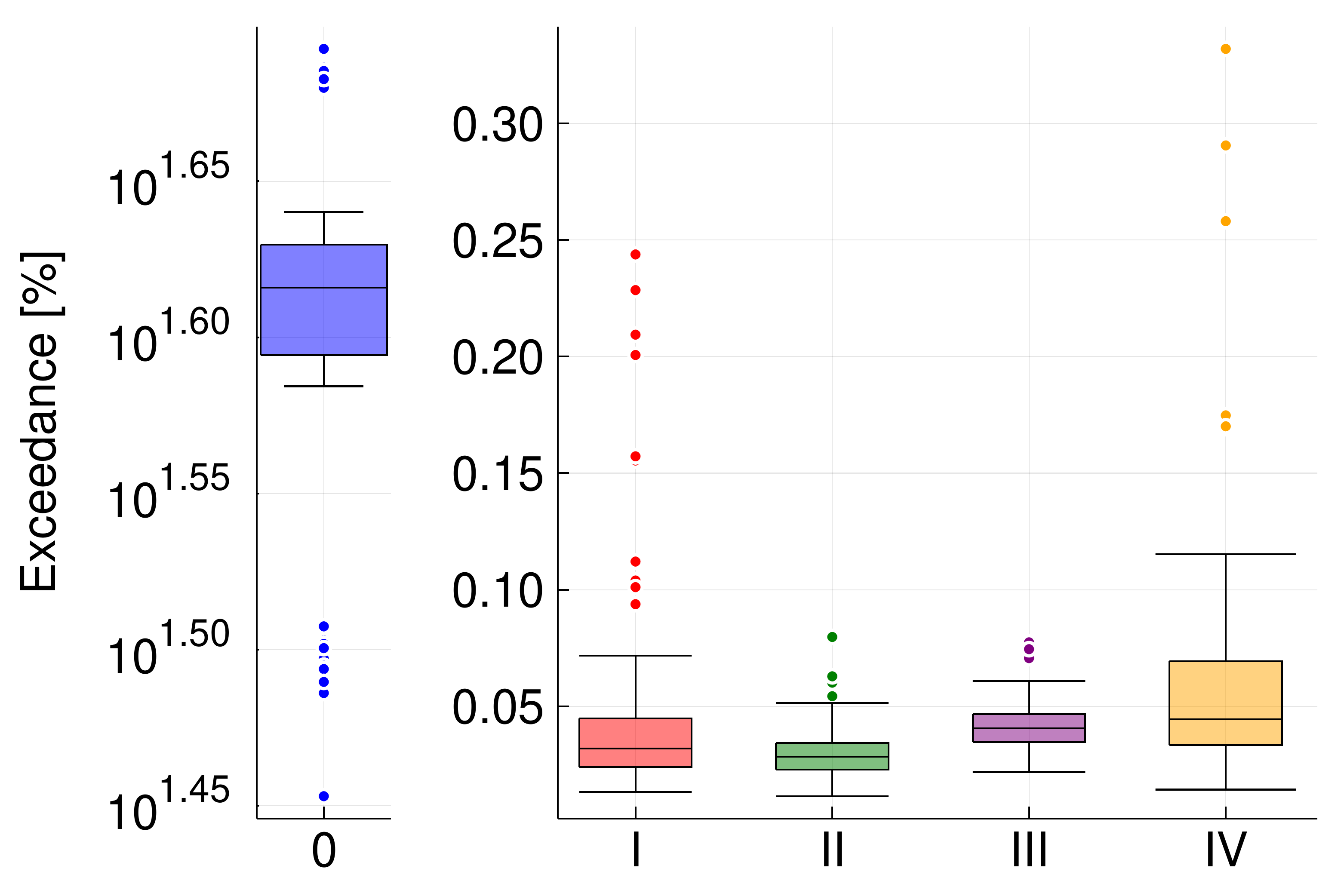}
\caption{The exceedance of the nodes for the various higher layers, see Tab.~\ref{tab:exp} for the scenario definitions.}
\label{fig:exceed}
\end{figure}

We first consider the maximum frequency deviation to see how far the system deviates from the desired frequency and the exceedance of the frequency. These observables show the quality of the control achieved by the lower layer in the presence of the various higher-layer topologies.
The first plot Fig.~\ref{fig:freq_dev} shows the distribution of $\omega_{top}$ across the grid ensemble. In each scenario, 
$\omega_{min}\ll\omega_{top}\ll\omega_{max}$, i.e. the lower-layer control objective Eq.~\eqref{eq:boundedf}
is always achieved. 
We see that the performance is very similar across the communication scenarios, it is slightly better for the no-ILC case ($\mathbf 0$) and the local ILC at all nodes without communication ($\mathbf I$). 
Fig.~\ref{fig:exceed} depicts the exceedance $exc_j$ across the grid ensemble for all nodes $j$. 
We can observe that adding a higher-layer control reduces the exceedance drastically ($\mathbf{I}$-$\mathbf{IV}$). The scenarios with communication ($\mathbf{II, IV}$) perform slightly better than without ($\mathbf{I, III}$). 
It is apparent from the observation of both $\omega_{top}$ and $exc_j$ that the addition of the higher layer
the amplitude of transient deviations but at the same time shortens their duration. This indicates that the 
ILC control suppresses demand fluctuations that drive the system out of the bounds given by $\Delta\omega$.

%Note that the frequency threshold for the exceedance is chosen very small here with 0.0005 rad/s. NoTab. is that the exceedance has a large tail of outliers in cases \textbf{I} and \textbf{IV}.
%We suspect that this is due to the fact that there is no (direct (\textbf{II}, \textbf{IV}) or indirect (\textbf{III})) spatial aggregation here, and that the ILC is overlearning the fluctuations at the nodes at which it is situated.

\begin{figure}[!htp]
\centering
\includegraphics[width=\columnwidth]{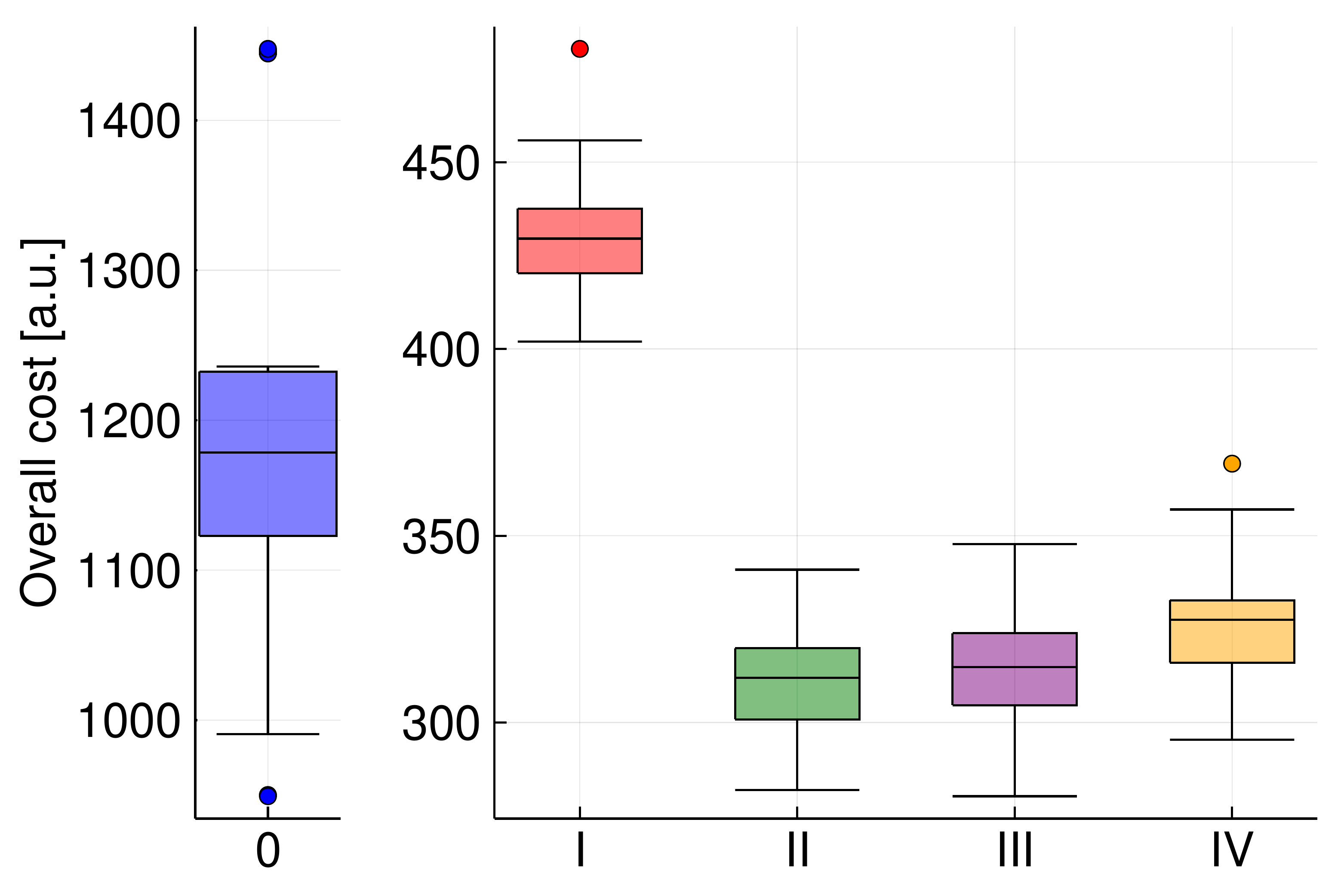}
\caption{The overall energy cost, sum over all nodes, logarithmic axis, 
see Tab.~\ref{tab:exp} for the scenario definitions.}
\label{fig:control_all_nodes}
\end{figure}

More interestingly, we can now consider the total system cost from Eq.~\eqref{eq:cost}  over the course of the last ten days of the simulations, given the various choices of higher-layer control. 
We chose $\lambda = 0.8$, i.e., instantaneous control energy being four times more expensive than energy at the day-ahead market.  The baseline scenario $\mathbf{0}$, with no ILC at all, in the left column of Fig.~\ref{fig:control_all_nodes}, gives us an idea of the total cost in the absence of dispatch. 
Providing dispatchable energy at every node in scenario $\mathbf{I}$ reduces the total cost by a factor of almost three with the parameters chosen, see the second column in Fig.~\ref{fig:control_all_nodes}. 

Turning now to the three main scenarios $\mathbf{II}$-$\mathbf{IV}$, we see that they all manage to reduce the cost even further compared to scenario $\mathbf{I}$. For those scenarios, we obtain a cost reduction factor of around four
compared to the non-ILC scenario \textbf{0}, consistent with our choice of $\lambda$ (${\lambda}/{1-\lambda}={0.8}/{0.2} = 4$).

Presumably, the increased cost in scenario $\mathbf{I}$ is due to conflicting actions of the decantralised controllers
that are eliminated by the communication infrastructure in the scenarios $\mathbf{II}$-$\mathbf{IV}$. 
If, however, the communication topology is not adapted to the unerlying physical network but
random, we also observe a slight cost increase, showing that random aggregation is generally not beneficial 
over placing control at a maximal independent set in scenario $\mathbf{II}$.

%Fig.~\ref{fig:control_ILC_nodes} and \ref{fig:control_nonILC_nodes} show the overall cost at ILC nodes and non-ILC nodes respectively. As we see, at the ILC nodes the no communication system $\mathbf{III}$ has a performance between the communication scenarios $\mathbf{II, IV}$ and is equal or better at non-ILC nodes. 

\begin{figure}[!htp]
\centering
\includegraphics[width=\columnwidth]{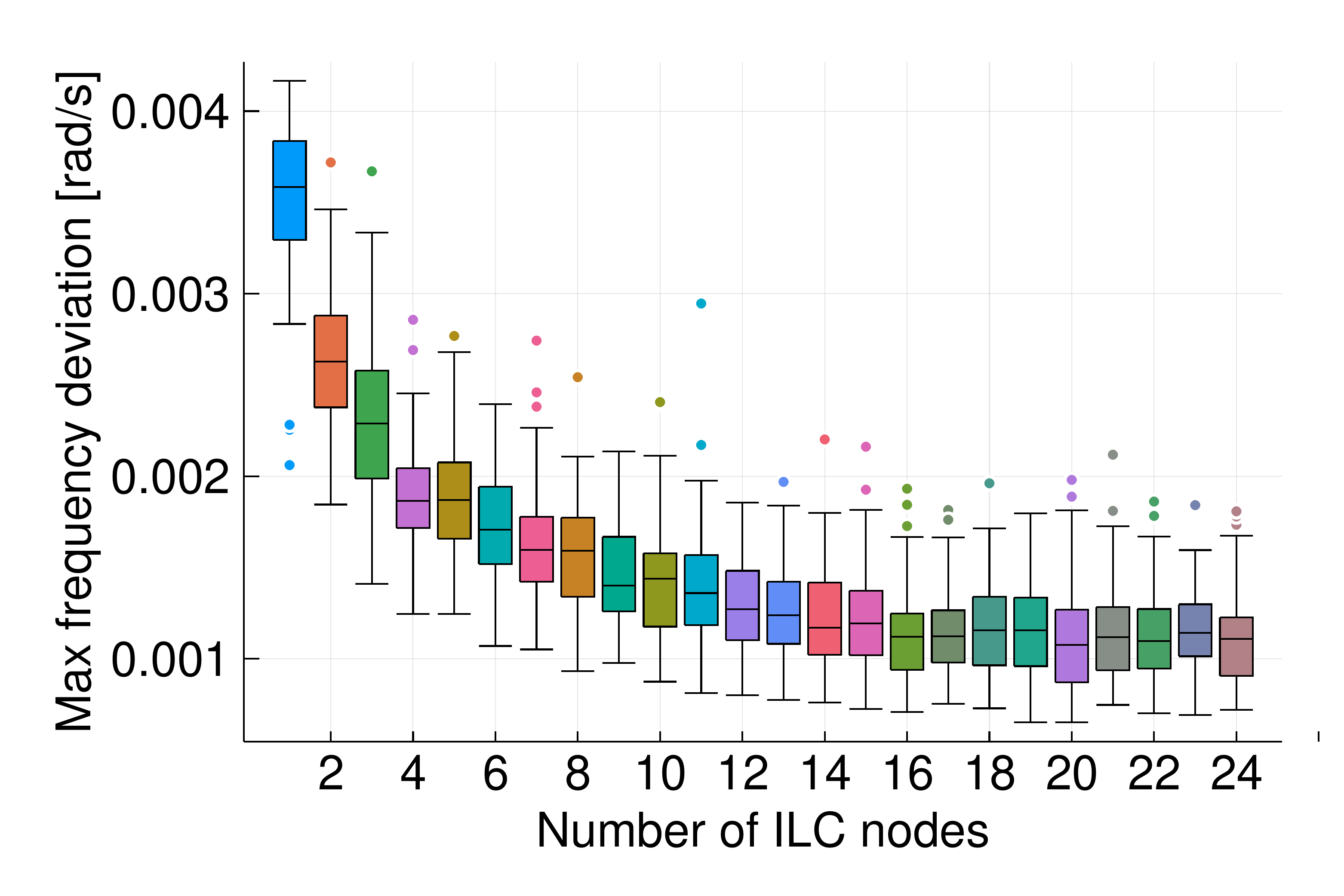}
\caption{Maximum frequency deviation for $\lambda = 0.8$ with variation of ILC nodes, 50 days simulated, last 10 days observed}
\label{fig:ilcnodes_max_freq}
\end{figure}

\begin{figure}[!htp]
\centering
\includegraphics[width=\columnwidth]{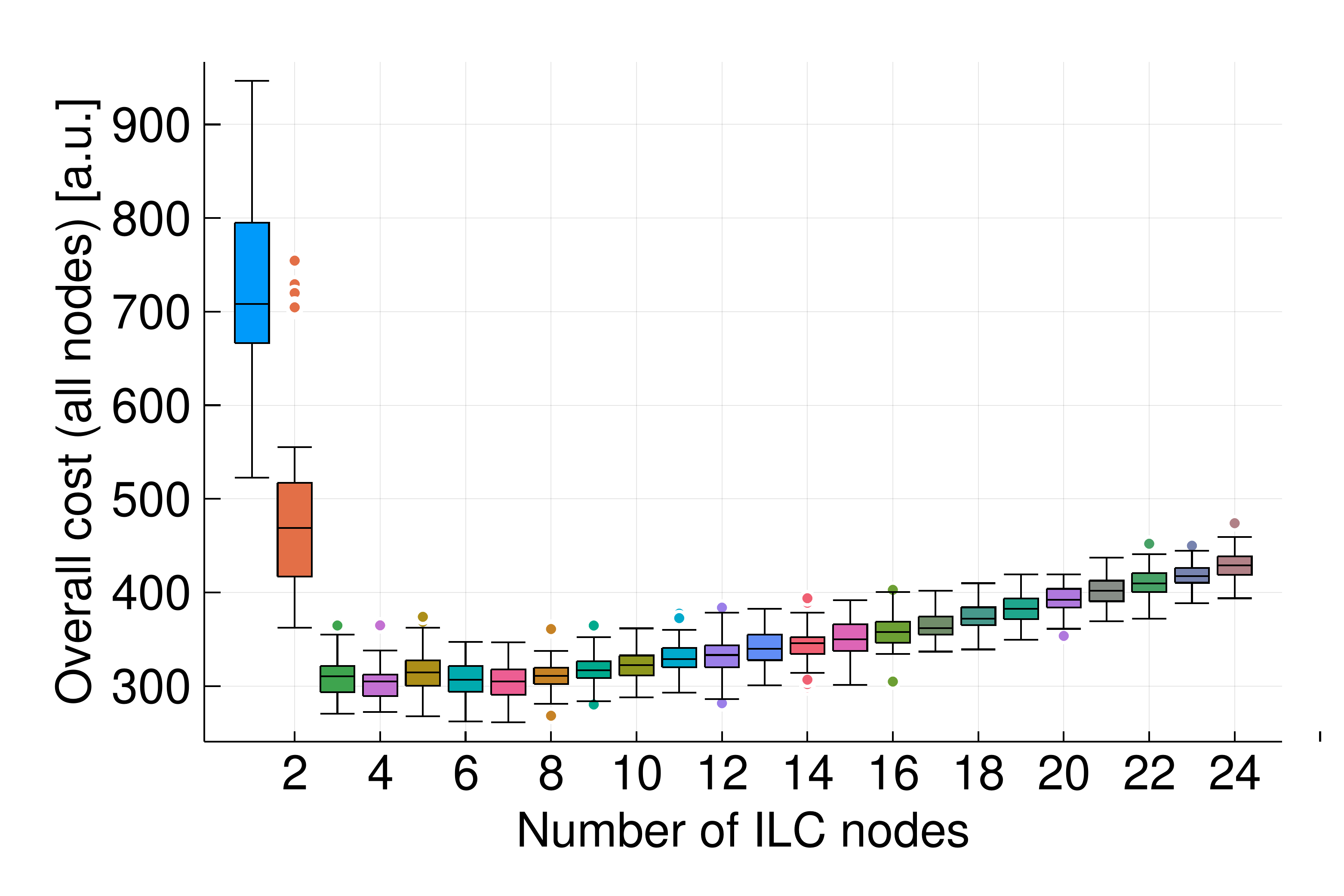}
\caption{Overall cost for $\lambda = 0.8$ with variation of ILC nodes, 50 days simulated, last 10 days observed}
\label{fig:ilcnodes}
\end{figure}

Another distinction between $\mathbf{I}$ and $\mathbf{II}$-$\mathbf{IV}$ is the number of nodes in 
the control set $\mathcal{N}^C$. To investigate the influence on the overall performance, 
we systematically varied the number of ILC nodes from 1 to $card(\mathcal{N})=24$. 
The controlled nodes were randomly drawn in each run and not connected to any other nodes in the communication layer similar to scenario $\mathbf{III}$. The results are two-fold.
The maximum frequency deviation in Fig.~\ref{fig:ilcnodes_max_freq} decreases monotonically, indicating
that the best performance is achieved with the highest control effort.
Contrarily, Fig.~\ref{fig:ilcnodes} shows that the system is cost-optimal when about one third of nodes
are equipped with ILC control. Whereas the significant improvement compared to $\mathbf{0}$ is 
evident for a small size of $\mathcal{N}^C$, a further distribution of control action across 
more nodes leads to slightly higher costs, still remaining well below the baseline scenario.
Interestingly, when $\mathcal{N}^C=\mathcal N$, the expected cost is higher than in scenario
$\mathbf{I}$ without communication. Concerning the cost difference between $\mathbf{I}$ and
$\mathbf{II}$-$\mathbf{IV}$ in Fig.~\ref{fig:control_all_nodes}, this experiment implies that the reduction is mainly achieved
by a smaller-sized control set whereas the addition of communication links actually increases
the costs.

\begin{table}[!htp]
\caption{Overview of the studied higher-layer topologies}
\label{tab:exp}
\vspace{2mm}
\begin{tabular}{|m{0.6cm}|m{3.25cm}|m{3.9cm}|}
\hline
\textbf{Exp.}&\textbf{Description} & \textbf{Communication graph}\\
\hline
\hline
$\mathbf{0}$ & no ILC & $\mathcal G^{C} $ is the empty graph  \\
\hline
$\mathbf{I}$ & local ILC at all nodes &
\specialcell{$\mathcal G^{C} = (\mathcal N, \emptyset)$,\\card($\mathcal{N}^C$) = card($\mathcal N$) }\\
\hline
$\mathbf{II}$ & ILC at nodes in a max. independent set in the network graph and averaged update with all neighbouring nodes &
\specialcell{$\mathcal G^{C} = (\mathcal N,\mathcal E^C)$,\\ $\mathcal E^C = \mathcal E \cap \left(\mathcal N^C \times \mathcal N^n\right)$,\\ $\mathcal N^C$ max. independent set} \\
\hline
$\mathbf{III}$ & local ILC at nodes in a max. independent set in the network graph & $\mathcal G^{C} = (\mathcal N, \emptyset)$, $\mathcal N^C$ max. independent set  \\
\hline
$\mathbf{IV}$ & ILC at 50 \% of the nodes  with averaged update with 3 random other nodes &
\specialcell{$\mathcal G^{C} = $($\mathcal N$, $\mathcal E^C$),\\ $\mathcal E^C =$\\ $ \{(i,j)|i \in \mathcal N^C, j\in \mathcal N\setminus \{i\} \}$,\\ card($\mathcal{N}^C$) = card($\mathcal N^n$)}\\
\hline
\hline
\end{tabular}
\end{table}

% \begin{figure}
% \centering
% \includegraphics[width=\columnwidth]{./figures/cost_boxplot_ilcnodes_lambda_0-8.pdf}
% \caption{The overall energy cost, sum over nodes with dispatchable 
% energy only (ILC nodes), logarithmic axis, see Tab.~\ref{tab:exp} for the scenario definitions.}
% \label{fig:control_ILC_nodes}
% \end{figure}

% \begin{figure}
% \centering
% \includegraphics[width=\columnwidth]{./figures/cost_boxplot_nonilcnodes_lambda_0-8.pdf}
% \caption{The overall energy cost, sum over nodes without dispatchable 
% energy only (non-ILC nodes), logarithmic axis, see Tab.~\ref{tab:exp} for the scenario definitions.}
% \label{fig:control_nonILC_nodes}
% \end{figure}

\section{Discussion, conclusion and way forward}
\label{s:discuss}

In this paper, we introduced a multiplex hierarchical model of power grids that covers timescales from seconds to days, and allows studying the interaction of energy management/tertiary control and self-organised primary and secondary control.

Remarkably, we find that a basic but natural communication and aggregation scheme in the higher layer does not 
improve the performance. 
The results indicate that the self-organised distributed control of the lower layer already carries sufficient information to learn the appropriate dispatch at those nodes that are dispatchable. 
Adding explicit communication does not visibly reduce the cost of the system while the performance 
with respect to the lower-layer control objective is expected to decrease slightly.

%This indicates that the self-organised behaviour of the lower layer carries important information for various higher layer control tasks in such a networked system.

Actually, there already exists implicit communication through the power flows managed by the primary and secondary control in each scenario. It seems that any additional communication that tries to aggregate the local deviations, needs to take this into account. Also a carefully designed Q-filter and learning matrix may improve the performance of additional communication. 

Clearly for more complex and interesting control objectives, that a realistic energy management system has to achieve, communication is necessary. 
However, our results show that even in these cases it might be worthwhile investigating the implicit communication already present in the system, and taking it into account since it leads to inputs taken from different parts of the system to be correlated, thus potentially causing an overcompensation.

More broadly, we saw that for the parametrisation of the decentralised control layer, 
probabilistic methods are a useful complement to analytic bounds. 
The selection of $k_{I,j}$ is chosen according to the analytical bounds in \cite[Corollary 4]{doerfler2018leakyintegrator} and to avoid a frequency deviation independently of the choice of $k_{P,j}$ (cp. Fig. \ref{fig:cp_kp1}).

Further, the presence of a higher layer with separate control objectives certainly has the potential to affect the performance of the lower layer. 
For any added ILC there is a large effect on the exceedance (Fig. \ref{fig:exceed}). 
This is not unexpected as the control fundamentally changes the nature of all nodes. 
Still, it indicates that it is interesting to further study the interactions of the layers in the future. 
In fact, the interaction between higher-layer energy management and the frequency dynamics of the power grid is an effect that is observed in real power grids, where trading intervals are very visible in the statistics of power grid frequency signals \cite{schafer2018non, anvari2019stochastic}. 
We expect that the type of model we have set up here is highly useful to study and reproduce some of these results without having to go to highly specific market models.

% \subsubsection{Does hierarchical control bring a benefit compared to no hierarchy?}
%We can state that even the lower layer control objectives can be improved by adding ILC and thus a hierarchical design. 
% It should be checked in future work if an added Q-filter and different amplitudes of the demand for different days will have a similar result. 
% \subsubsection{ Does the high-level controller with ILC need communication?}
% We can not confirm that adding communication improves the performance, in contrary, in our case it is neutral to worse. In future research, the ILC needs to be adapted in more detail to be able to make this statement in a more general way.
%\\ \texttt{Later if we have time and space: Can the Iterative Learning Control (ILC) recreate a nice post-fault states?}

%\section{Conclusion and way forward}
%The results show that for this setting, local ILC at all nodes is the most beneficial way to save control energy on the lower layer. It can also be applied to the nodes in a vertex cover of the network graph to have some effect. Neighboring or random communication seem not to bring a benefit. 

The general setup we have chosen can serve as a wide ranging basis for the study of the interaction of dynamics and dispatch constraints. 
The overall model can easily be extended to include local limits on available storage and ramping times. 
Adding individual and time-varying prices for various forms of energy to give the ILC a more realistic target function to optimise is also straightforward. 

Further, a more sophisticated higher layer can use the distributed controllers to achieve more challenging goals than merely minimising price, or prioritising one type of energy over the other, i.e., in future work, we may consider, e.g., the set $[P^{ILC}_{j}(t), T_j(t), k_{p,j}(t), k_{I,j}(t)]$ as an input to the ILC.

% In order to address the drawbacks of the control, future work should investigate if  incorporating the lower level control parameters being learned by the ILC is beneficial for the performance. In addition, it should be explored if more sophisticated design procedures or different ILC designs - linear and nonlinear - show similar results when communication is added. 

% In a next step, the learning can be expanded to learn the lower layer control parameters. In addition, different node types should be added to the model.

Finally, we note that the type of system we introduced here can be of independent interest in the context of theoretical physics. 
For example, \citet{Nicosia2014a} analyse the coupling between different dynamics in a multiplex network, i.e.,  between a network of Kuramoto oscillators and a random walk. 
Under certain conditions, the coupling between the layers then induces spontaneous explosive synchronisation transitions. 
Since we study the synchronisation of Kuramoto	oscillators with inertia (Eq.~\eqref{eq:mn_plant}) and use proportional ILC that is also linear, this is mathematically similar to our model. 
Thus it would be interesting under which conditions models like ours can exhibit such properties as well. 

\section{Software}

All code was written in Julia and is available on request or on the first authors github\footnote{Repository controlled-multi-timescale-powergrid at \url{https://github.com/strangeli/controlled-multi-timescale-powergrid}}. The simulations were performed using the DifferentialEquations.jl package\cite{rackauckas2017differentialequations} and the Rodas4p solver\cite{wanner1996solving}.

\begin{acknowledgments}
This work was funded by the Deutsche Forschungsgemeinschaft (DFG, German Research Foundation) – KU 837/39-1 / RA 516/13-1.
\end{acknowledgments}

\section*{References}
\bibliography{literature}

\begin{thebibliography}{59}%
\makeatletter
\providecommand \@ifxundefined [1]{%
 \@ifx{#1\undefined}
}%
\providecommand \@ifnum [1]{%
 \ifnum #1\expandafter \@firstoftwo
 \else \expandafter \@secondoftwo
 \fi
}%
\providecommand \@ifx [1]{%
 \ifx #1\expandafter \@firstoftwo
 \else \expandafter \@secondoftwo
 \fi
}%
\providecommand \natexlab [1]{#1}%
\providecommand \enquote  [1]{``#1''}%
\providecommand \bibnamefont  [1]{#1}%
\providecommand \bibfnamefont [1]{#1}%
\providecommand \citenamefont [1]{#1}%
\providecommand \href@noop [0]{\@secondoftwo}%
\providecommand \href [0]{\begingroup \@sanitize@url \@href}%
\providecommand \@href[1]{\@@startlink{#1}\@@href}%
\providecommand \@@href[1]{\endgroup#1\@@endlink}%
\providecommand \@sanitize@url [0]{\catcode `\\12\catcode `\$12\catcode
  `\&12\catcode `\#12\catcode `\^12\catcode `\_12\catcode `\%12\relax}%
\providecommand \@@startlink[1]{}%
\providecommand \@@endlink[0]{}%
\providecommand \url  [0]{\begingroup\@sanitize@url \@url }%
\providecommand \@url [1]{\endgroup\@href {#1}{\urlprefix }}%
\providecommand \urlprefix  [0]{URL }%
\providecommand \Eprint [0]{\href }%
\providecommand \doibase [0]{http://dx.doi.org/}%
\providecommand \selectlanguage [0]{\@gobble}%
\providecommand \bibinfo  [0]{\@secondoftwo}%
\providecommand \bibfield  [0]{\@secondoftwo}%
\providecommand \translation [1]{[#1]}%
\providecommand \BibitemOpen [0]{}%
\providecommand \bibitemStop [0]{}%
\providecommand \bibitemNoStop [0]{.\EOS\space}%
\providecommand \EOS [0]{\spacefactor3000\relax}%
\providecommand \BibitemShut  [1]{\csname bibitem#1\endcsname}%
\let\auto@bib@innerbib\@empty
%</preamble>
\bibitem [{\citenamefont {Guerrero}\ \emph {et~al.}(2010)\citenamefont
  {Guerrero}, \citenamefont {Vasquez}, \citenamefont {Matas}, \citenamefont
  {De~Vicu{\~n}a},\ and\ \citenamefont {Castilla}}]{Guerrero2010}%
  \BibitemOpen
  \bibfield  {author} {\bibinfo {author} {\bibfnamefont {J.~M.}\ \bibnamefont
  {Guerrero}}, \bibinfo {author} {\bibfnamefont {J.~C.}\ \bibnamefont
  {Vasquez}}, \bibinfo {author} {\bibfnamefont {J.}~\bibnamefont {Matas}},
  \bibinfo {author} {\bibfnamefont {L.~G.}\ \bibnamefont {De~Vicu{\~n}a}}, \
  and\ \bibinfo {author} {\bibfnamefont {M.}~\bibnamefont {Castilla}},\
  }\bibfield  {title} {\enquote {\bibinfo {title} {Hierarchical control of
  droop-controlled {AC} and {DC} microgrids---a general approach toward
  standardization},}\ }\href@noop {} {\bibfield  {journal} {\bibinfo  {journal}
  {IEEE Transactions on Industrial Electronics}\ }\textbf {\bibinfo {volume}
  {58}},\ \bibinfo {pages} {158--172} (\bibinfo {year} {2010})}\BibitemShut
  {NoStop}%
\bibitem [{\citenamefont {Schiffer}\ \emph {et~al.}(2015)\citenamefont
  {Schiffer}, \citenamefont {Zonetti}, \citenamefont {Ortega}, \citenamefont
  {Stankovicc}, \citenamefont {Sezid},\ and\ \citenamefont
  {Raisch}}]{schiffer2015modeling}%
  \BibitemOpen
  \bibfield  {author} {\bibinfo {author} {\bibfnamefont {J.}~\bibnamefont
  {Schiffer}}, \bibinfo {author} {\bibfnamefont {D.}~\bibnamefont {Zonetti}},
  \bibinfo {author} {\bibfnamefont {R.}~\bibnamefont {Ortega}}, \bibinfo
  {author} {\bibfnamefont {A.}~\bibnamefont {Stankovicc}}, \bibinfo {author}
  {\bibfnamefont {T.}~\bibnamefont {Sezid}}, \ and\ \bibinfo {author}
  {\bibfnamefont {J.}~\bibnamefont {Raisch}},\ }\bibfield  {title} {\enquote
  {\bibinfo {title} {Modeling of microgrids---from fundamental physics to
  phasors and voltage sources},}\ }\href@noop {} {\bibfield  {journal}
  {\bibinfo  {journal} {Automatica}\ ,\ \bibinfo {pages} {1--15}} (\bibinfo
  {year} {2015})}\BibitemShut {NoStop}%
\bibitem [{\citenamefont {Bidram}\ and\ \citenamefont
  {Davoudi}(2012)}]{bidram2012hierarchical}%
  \BibitemOpen
  \bibfield  {author} {\bibinfo {author} {\bibfnamefont {A.}~\bibnamefont
  {Bidram}}\ and\ \bibinfo {author} {\bibfnamefont {A.}~\bibnamefont
  {Davoudi}},\ }\bibfield  {title} {\enquote {\bibinfo {title} {Hierarchical
  structure of microgrids control system},}\ }\href@noop {} {\bibfield
  {journal} {\bibinfo  {journal} {IEEE Transactions on Smart Grid}\ }\textbf
  {\bibinfo {volume} {3}},\ \bibinfo {pages} {1963--1976} (\bibinfo {year}
  {2012})}\BibitemShut {NoStop}%
\bibitem [{\citenamefont {D{\"o}rfler}, \citenamefont {Simpson-Porco},\ and\
  \citenamefont {Bullo}(2014)}]{dorfler2014plug}%
  \BibitemOpen
  \bibfield  {author} {\bibinfo {author} {\bibfnamefont {F.}~\bibnamefont
  {D{\"o}rfler}}, \bibinfo {author} {\bibfnamefont {J.~W.}\ \bibnamefont
  {Simpson-Porco}}, \ and\ \bibinfo {author} {\bibfnamefont {F.}~\bibnamefont
  {Bullo}},\ }\bibfield  {title} {\enquote {\bibinfo {title} {Plug-and-play
  control and optimization in microgrids},}\ }in\ \href@noop {} {\emph
  {\bibinfo {booktitle} {53rd IEEE Conference on Decision and Control}}}\
  (\bibinfo {organization} {IEEE},\ \bibinfo {year} {2014})\ pp.\ \bibinfo
  {pages} {211--216}\BibitemShut {NoStop}%
\bibitem [{\citenamefont {Olivares}\ \emph {et~al.}(2014)\citenamefont
  {Olivares}, \citenamefont {Mehrizi-Sani}, \citenamefont {Etemadi},
  \citenamefont {Ca{\~n}izares}, \citenamefont {Iravani}, \citenamefont
  {Kazerani}, \citenamefont {Hajimiragha}, \citenamefont {Gomis-Bellmunt},
  \citenamefont {Saeedifard}, \citenamefont {Palma-Behnke} \emph
  {et~al.}}]{olivares2014trends}%
  \BibitemOpen
  \bibfield  {author} {\bibinfo {author} {\bibfnamefont {D.~E.}\ \bibnamefont
  {Olivares}}, \bibinfo {author} {\bibfnamefont {A.}~\bibnamefont
  {Mehrizi-Sani}}, \bibinfo {author} {\bibfnamefont {A.~H.}\ \bibnamefont
  {Etemadi}}, \bibinfo {author} {\bibfnamefont {C.~A.}\ \bibnamefont
  {Ca{\~n}izares}}, \bibinfo {author} {\bibfnamefont {R.}~\bibnamefont
  {Iravani}}, \bibinfo {author} {\bibfnamefont {M.}~\bibnamefont {Kazerani}},
  \bibinfo {author} {\bibfnamefont {A.~H.}\ \bibnamefont {Hajimiragha}},
  \bibinfo {author} {\bibfnamefont {O.}~\bibnamefont {Gomis-Bellmunt}},
  \bibinfo {author} {\bibfnamefont {M.}~\bibnamefont {Saeedifard}}, \bibinfo
  {author} {\bibfnamefont {R.}~\bibnamefont {Palma-Behnke}},  \emph {et~al.},\
  }\bibfield  {title} {\enquote {\bibinfo {title} {Trends in microgrid
  control},}\ }\href@noop {} {\bibfield  {journal} {\bibinfo  {journal} {IEEE
  Transactions on Smart Grid}\ }\textbf {\bibinfo {volume} {5}},\ \bibinfo
  {pages} {1905--1919} (\bibinfo {year} {2014})}\BibitemShut {NoStop}%
\bibitem [{\citenamefont {Aamir}, \citenamefont {Kalwar},\ and\ \citenamefont
  {Mekhilef}(2016)}]{Aamir2016}%
  \BibitemOpen
  \bibfield  {author} {\bibinfo {author} {\bibfnamefont {M.}~\bibnamefont
  {Aamir}}, \bibinfo {author} {\bibfnamefont {K.~A.}\ \bibnamefont {Kalwar}}, \
  and\ \bibinfo {author} {\bibfnamefont {S.}~\bibnamefont {Mekhilef}},\
  }\bibfield  {title} {\enquote {\bibinfo {title} {Uninterruptible power supply
  (ups) system},}\ }\href@noop {} {\bibfield  {journal} {\bibinfo  {journal}
  {Renewable and Sustainable Energy Reviews}\ }\textbf {\bibinfo {volume}
  {58}},\ \bibinfo {pages} {1395--1410} (\bibinfo {year} {2016})}\BibitemShut
  {NoStop}%
\bibitem [{\citenamefont {Xin}\ \emph {et~al.}(2015)\citenamefont {Xin},
  \citenamefont {Zhao}, \citenamefont {Zhang}, \citenamefont {Wang},
  \citenamefont {Wong},\ and\ \citenamefont {Wei}}]{xin2015decentralized}%
  \BibitemOpen
  \bibfield  {author} {\bibinfo {author} {\bibfnamefont {H.}~\bibnamefont
  {Xin}}, \bibinfo {author} {\bibfnamefont {R.}~\bibnamefont {Zhao}}, \bibinfo
  {author} {\bibfnamefont {L.}~\bibnamefont {Zhang}}, \bibinfo {author}
  {\bibfnamefont {Z.}~\bibnamefont {Wang}}, \bibinfo {author} {\bibfnamefont
  {K.~P.}\ \bibnamefont {Wong}}, \ and\ \bibinfo {author} {\bibfnamefont
  {W.}~\bibnamefont {Wei}},\ }\bibfield  {title} {\enquote {\bibinfo {title} {A
  decentralized hierarchical control structure and self-optimizing control
  strategy for {FP} type {DGs} in islanded microgrids},}\ }\href@noop {}
  {\bibfield  {journal} {\bibinfo  {journal} {IEEE Transactions on Smart Grid}\
  }\textbf {\bibinfo {volume} {7}},\ \bibinfo {pages} {3--5} (\bibinfo {year}
  {2015})}\BibitemShut {NoStop}%
\bibitem [{\citenamefont {Han}\ \emph {et~al.}(2016)\citenamefont {Han},
  \citenamefont {Li}, \citenamefont {Shen}, \citenamefont {Coelho},\ and\
  \citenamefont {Guerrero}}]{han2016review}%
  \BibitemOpen
  \bibfield  {author} {\bibinfo {author} {\bibfnamefont {Y.}~\bibnamefont
  {Han}}, \bibinfo {author} {\bibfnamefont {H.}~\bibnamefont {Li}}, \bibinfo
  {author} {\bibfnamefont {P.}~\bibnamefont {Shen}}, \bibinfo {author}
  {\bibfnamefont {E.~A.~A.}\ \bibnamefont {Coelho}}, \ and\ \bibinfo {author}
  {\bibfnamefont {J.~M.}\ \bibnamefont {Guerrero}},\ }\bibfield  {title}
  {\enquote {\bibinfo {title} {Review of active and reactive power sharing
  strategies in hierarchical controlled microgrids},}\ }\href@noop {}
  {\bibfield  {journal} {\bibinfo  {journal} {IEEE Transactions on Power
  Electronics}\ }\textbf {\bibinfo {volume} {32}},\ \bibinfo {pages}
  {2427--2451} (\bibinfo {year} {2016})}\BibitemShut {NoStop}%
\bibitem [{\citenamefont {Li}\ \emph {et~al.}(2017)\citenamefont {Li},
  \citenamefont {Zang}, \citenamefont {Zeng}, \citenamefont {Yu},\ and\
  \citenamefont {Li}}]{li2017fully}%
  \BibitemOpen
  \bibfield  {author} {\bibinfo {author} {\bibfnamefont {Z.}~\bibnamefont
  {Li}}, \bibinfo {author} {\bibfnamefont {C.}~\bibnamefont {Zang}}, \bibinfo
  {author} {\bibfnamefont {P.}~\bibnamefont {Zeng}}, \bibinfo {author}
  {\bibfnamefont {H.}~\bibnamefont {Yu}}, \ and\ \bibinfo {author}
  {\bibfnamefont {S.}~\bibnamefont {Li}},\ }\bibfield  {title} {\enquote
  {\bibinfo {title} {Fully distributed hierarchical control of parallel
  grid-supporting inverters in islanded {AC} microgrids},}\ }\href@noop {}
  {\bibfield  {journal} {\bibinfo  {journal} {IEEE Transactions on Industrial
  Informatics}\ }\textbf {\bibinfo {volume} {14}},\ \bibinfo {pages} {679--690}
  (\bibinfo {year} {2017})}\BibitemShut {NoStop}%
\bibitem [{\citenamefont {Gao}\ \emph {et~al.}(2012)\citenamefont {Gao},
  \citenamefont {Buldyrev}, \citenamefont {Stanley},\ and\ \citenamefont
  {Havlin}}]{Gao2011a}%
  \BibitemOpen
  \bibfield  {author} {\bibinfo {author} {\bibfnamefont {J.}~\bibnamefont
  {Gao}}, \bibinfo {author} {\bibfnamefont {S.~V.}\ \bibnamefont {Buldyrev}},
  \bibinfo {author} {\bibfnamefont {H.~E.}\ \bibnamefont {Stanley}}, \ and\
  \bibinfo {author} {\bibfnamefont {S.}~\bibnamefont {Havlin}},\ }\bibfield
  {title} {\enquote {\bibinfo {title} {{Networks formed from interdependent
  networks}},}\ }\href@noop {} {\bibfield  {journal} {\bibinfo  {journal}
  {Nature Physics}\ }\textbf {\bibinfo {volume} {8}},\ \bibinfo {pages}
  {40--48} (\bibinfo {year} {2012})}\BibitemShut {NoStop}%
\bibitem [{\citenamefont {Gale}, \citenamefont {Kariv},\ and\ \citenamefont
  {Systems}(2014)}]{Gale2014}%
  \BibitemOpen
  \bibfield  {author} {\bibinfo {author} {\bibfnamefont {D.~M.}\ \bibnamefont
  {Gale}}, \bibinfo {author} {\bibfnamefont {S.}~\bibnamefont {Kariv}}, \ and\
  \bibinfo {author} {\bibfnamefont {U.~C.}\ \bibnamefont {Systems}},\
  }\href@noop {} {\emph {\bibinfo {title} {American Economic Review}}},\ edited
  by\ \bibinfo {editor} {\bibfnamefont {G.}~\bibnamefont {D'Agostino}}\ and\
  \bibinfo {editor} {\bibfnamefont {A.}~\bibnamefont {Scala}},\ \bibinfo
  {series} {Understanding Complex Systems}, Vol.~\bibinfo {volume} {97}\
  (\bibinfo  {publisher} {Springer International Publishing},\ \bibinfo
  {address} {Cham},\ \bibinfo {year} {2014})\ pp.\ \bibinfo {pages}
  {99--103}\BibitemShut {NoStop}%
\bibitem [{\citenamefont {Kivel{\"{a}}}\ \emph {et~al.}(2014)\citenamefont
  {Kivel{\"{a}}}, \citenamefont {Arenas}, \citenamefont {Barthelemy},
  \citenamefont {Gleeson}, \citenamefont {Moreno},\ and\ \citenamefont
  {Porter}}]{Kivela2014}%
  \BibitemOpen
  \bibfield  {author} {\bibinfo {author} {\bibfnamefont {M.}~\bibnamefont
  {Kivel{\"{a}}}}, \bibinfo {author} {\bibfnamefont {A.}~\bibnamefont
  {Arenas}}, \bibinfo {author} {\bibfnamefont {M.}~\bibnamefont {Barthelemy}},
  \bibinfo {author} {\bibfnamefont {J.~P.}\ \bibnamefont {Gleeson}}, \bibinfo
  {author} {\bibfnamefont {Y.}~\bibnamefont {Moreno}}, \ and\ \bibinfo {author}
  {\bibfnamefont {M.~A.}\ \bibnamefont {Porter}},\ }\bibfield  {title}
  {\enquote {\bibinfo {title} {{Multilayer networks}},}\ }\href@noop {}
  {\bibfield  {journal} {\bibinfo  {journal} {Journal of Complex Networks}\
  }\textbf {\bibinfo {volume} {2}},\ \bibinfo {pages} {203--271} (\bibinfo
  {year} {2014})}\BibitemShut {NoStop}%
\bibitem [{\citenamefont {{De Domenico}}\ \emph {et~al.}(2014)\citenamefont
  {{De Domenico}}, \citenamefont {Sol{\'{e}}-Ribalta}, \citenamefont {Cozzo},
  \citenamefont {Kivel{\"{a}}}, \citenamefont {Moreno}, \citenamefont {Porter},
  \citenamefont {G{\'{o}}mez},\ and\ \citenamefont {Arenas}}]{DeDomenico2013a}%
  \BibitemOpen
  \bibfield  {author} {\bibinfo {author} {\bibfnamefont {M.}~\bibnamefont {{De
  Domenico}}}, \bibinfo {author} {\bibfnamefont {A.}~\bibnamefont
  {Sol{\'{e}}-Ribalta}}, \bibinfo {author} {\bibfnamefont {E.}~\bibnamefont
  {Cozzo}}, \bibinfo {author} {\bibfnamefont {M.}~\bibnamefont {Kivel{\"{a}}}},
  \bibinfo {author} {\bibfnamefont {Y.}~\bibnamefont {Moreno}}, \bibinfo
  {author} {\bibfnamefont {M.~A.}\ \bibnamefont {Porter}}, \bibinfo {author}
  {\bibfnamefont {S.}~\bibnamefont {G{\'{o}}mez}}, \ and\ \bibinfo {author}
  {\bibfnamefont {A.}~\bibnamefont {Arenas}},\ }\bibfield  {title} {\enquote
  {\bibinfo {title} {{Mathematical formulation of multilayer networks}},}\
  }\href@noop {} {\bibfield  {journal} {\bibinfo  {journal} {Physical Review
  X}\ }\textbf {\bibinfo {volume} {3}},\ \bibinfo {pages} {1--15} (\bibinfo
  {year} {2014})}\BibitemShut {NoStop}%
\bibitem [{\citenamefont {Nicosia}\ \emph {et~al.}(2013)\citenamefont
  {Nicosia}, \citenamefont {Bianconi}, \citenamefont {Latora},\ and\
  \citenamefont {Barthelemy}}]{Nicosia2013}%
  \BibitemOpen
  \bibfield  {author} {\bibinfo {author} {\bibfnamefont {V.}~\bibnamefont
  {Nicosia}}, \bibinfo {author} {\bibfnamefont {G.}~\bibnamefont {Bianconi}},
  \bibinfo {author} {\bibfnamefont {V.}~\bibnamefont {Latora}}, \ and\ \bibinfo
  {author} {\bibfnamefont {M.}~\bibnamefont {Barthelemy}},\ }\bibfield  {title}
  {\enquote {\bibinfo {title} {{Growing multiplex networks}},}\ }\href@noop {}
  {\bibfield  {journal} {\bibinfo  {journal} {Physical Review Letters}\
  }\textbf {\bibinfo {volume} {111}},\ \bibinfo {pages} {1--5} (\bibinfo {year}
  {2013})}\BibitemShut {NoStop}%
\bibitem [{\citenamefont {Boccaletti}\ \emph {et~al.}(2014)\citenamefont
  {Boccaletti}, \citenamefont {Bianconi}, \citenamefont {Criado}, \citenamefont
  {del Genio}, \citenamefont {G{\'{o}}mez-Garde{\~{n}}es}, \citenamefont
  {Romance}, \citenamefont {Sendi{\~{n}}a-Nadal}, \citenamefont {Wang},\ and\
  \citenamefont {Zanin}}]{Boccaletti2014}%
  \BibitemOpen
  \bibfield  {author} {\bibinfo {author} {\bibfnamefont {S.}~\bibnamefont
  {Boccaletti}}, \bibinfo {author} {\bibfnamefont {G.}~\bibnamefont
  {Bianconi}}, \bibinfo {author} {\bibfnamefont {R.}~\bibnamefont {Criado}},
  \bibinfo {author} {\bibfnamefont {C.~I.}\ \bibnamefont {del Genio}}, \bibinfo
  {author} {\bibfnamefont {J.}~\bibnamefont {G{\'{o}}mez-Garde{\~{n}}es}},
  \bibinfo {author} {\bibfnamefont {M.}~\bibnamefont {Romance}}, \bibinfo
  {author} {\bibfnamefont {I.}~\bibnamefont {Sendi{\~{n}}a-Nadal}}, \bibinfo
  {author} {\bibfnamefont {Z.}~\bibnamefont {Wang}}, \ and\ \bibinfo {author}
  {\bibfnamefont {M.}~\bibnamefont {Zanin}},\ }\bibfield  {title} {\enquote
  {\bibinfo {title} {{The structure and dynamics of multilayer networks}},}\
  }\href@noop {} {\bibfield  {journal} {\bibinfo  {journal} {Physical Reports}\
  }\textbf {\bibinfo {volume} {544}},\ \bibinfo {pages} {1--122} (\bibinfo
  {year} {2014})}\BibitemShut {NoStop}%
\bibitem [{\citenamefont {Menck}\ \emph {et~al.}(2013)\citenamefont {Menck},
  \citenamefont {Heitzig}, \citenamefont {Marwan},\ and\ \citenamefont
  {Kurths}}]{menck2013basin}%
  \BibitemOpen
  \bibfield  {author} {\bibinfo {author} {\bibfnamefont {P.~J.}\ \bibnamefont
  {Menck}}, \bibinfo {author} {\bibfnamefont {J.}~\bibnamefont {Heitzig}},
  \bibinfo {author} {\bibfnamefont {N.}~\bibnamefont {Marwan}}, \ and\ \bibinfo
  {author} {\bibfnamefont {J.}~\bibnamefont {Kurths}},\ }\bibfield  {title}
  {\enquote {\bibinfo {title} {How basin stability complements the
  linear-stability paradigm},}\ }\href@noop {} {\bibfield  {journal} {\bibinfo
  {journal} {Nature Physics}\ }\textbf {\bibinfo {volume} {9}},\ \bibinfo
  {pages} {89} (\bibinfo {year} {2013})}\BibitemShut {NoStop}%
\bibitem [{\citenamefont {Menck}\ \emph {et~al.}(2014)\citenamefont {Menck},
  \citenamefont {Heitzig}, \citenamefont {Kurths},\ and\ \citenamefont
  {Schellnhuber}}]{menck2014dead}%
  \BibitemOpen
  \bibfield  {author} {\bibinfo {author} {\bibfnamefont {P.~J.}\ \bibnamefont
  {Menck}}, \bibinfo {author} {\bibfnamefont {J.}~\bibnamefont {Heitzig}},
  \bibinfo {author} {\bibfnamefont {J.}~\bibnamefont {Kurths}}, \ and\ \bibinfo
  {author} {\bibfnamefont {H.~J.}\ \bibnamefont {Schellnhuber}},\ }\bibfield
  {title} {\enquote {\bibinfo {title} {How dead ends undermine power grid
  stability},}\ }\href@noop {} {\bibfield  {journal} {\bibinfo  {journal}
  {Nature Communications}\ }\textbf {\bibinfo {volume} {5}},\ \bibinfo {pages}
  {3969} (\bibinfo {year} {2014})}\BibitemShut {NoStop}%
\bibitem [{\citenamefont {Schultz}, \citenamefont {Heitzig},\ and\
  \citenamefont {Kurths}(2014)}]{schultz2014detours}%
  \BibitemOpen
  \bibfield  {author} {\bibinfo {author} {\bibfnamefont {P.}~\bibnamefont
  {Schultz}}, \bibinfo {author} {\bibfnamefont {J.}~\bibnamefont {Heitzig}}, \
  and\ \bibinfo {author} {\bibfnamefont {J.}~\bibnamefont {Kurths}},\
  }\bibfield  {title} {\enquote {\bibinfo {title} {Detours around basin
  stability in power networks},}\ }\href@noop {} {\bibfield  {journal}
  {\bibinfo  {journal} {New Journal of Physics}\ }\textbf {\bibinfo {volume}
  {16}},\ \bibinfo {pages} {125001} (\bibinfo {year} {2014})}\BibitemShut
  {NoStop}%
\bibitem [{\citenamefont {Hellmann}\ \emph {et~al.}(2016)\citenamefont
  {Hellmann}, \citenamefont {Schultz}, \citenamefont {Grabow}, \citenamefont
  {Heitzig},\ and\ \citenamefont {Kurths}}]{hellmann2016survivability}%
  \BibitemOpen
  \bibfield  {author} {\bibinfo {author} {\bibfnamefont {F.}~\bibnamefont
  {Hellmann}}, \bibinfo {author} {\bibfnamefont {P.}~\bibnamefont {Schultz}},
  \bibinfo {author} {\bibfnamefont {C.}~\bibnamefont {Grabow}}, \bibinfo
  {author} {\bibfnamefont {J.}~\bibnamefont {Heitzig}}, \ and\ \bibinfo
  {author} {\bibfnamefont {J.}~\bibnamefont {Kurths}},\ }\bibfield  {title}
  {\enquote {\bibinfo {title} {Survivability of deterministic dynamical
  systems},}\ }\href@noop {} {\bibfield  {journal} {\bibinfo  {journal}
  {Scientific Reports}\ }\textbf {\bibinfo {volume} {6}},\ \bibinfo {pages}
  {29654} (\bibinfo {year} {2016})}\BibitemShut {NoStop}%
\bibitem [{\citenamefont {Schultz}\ \emph {et~al.}(2018)\citenamefont
  {Schultz}, \citenamefont {Hellmann}, \citenamefont {Webster},\ and\
  \citenamefont {Kurths}}]{schultz2018bounding}%
  \BibitemOpen
  \bibfield  {author} {\bibinfo {author} {\bibfnamefont {P.}~\bibnamefont
  {Schultz}}, \bibinfo {author} {\bibfnamefont {F.}~\bibnamefont {Hellmann}},
  \bibinfo {author} {\bibfnamefont {K.~N.}\ \bibnamefont {Webster}}, \ and\
  \bibinfo {author} {\bibfnamefont {J.}~\bibnamefont {Kurths}},\ }\bibfield
  {title} {\enquote {\bibinfo {title} {Bounding the first exit from the basin:
  Independence times and finite-time basin stability},}\ }\href@noop {}
  {\bibfield  {journal} {\bibinfo  {journal} {Chaos: An Interdisciplinary
  Journal of Nonlinear Science}\ }\textbf {\bibinfo {volume} {28}},\ \bibinfo
  {pages} {043102} (\bibinfo {year} {2018})}\BibitemShut {NoStop}%
\bibitem [{\citenamefont {Lindner}\ and\ \citenamefont
  {Hellmann}(2019)}]{lindner2019stochastic}%
  \BibitemOpen
  \bibfield  {author} {\bibinfo {author} {\bibfnamefont {M.}~\bibnamefont
  {Lindner}}\ and\ \bibinfo {author} {\bibfnamefont {F.}~\bibnamefont
  {Hellmann}},\ }\bibfield  {title} {\enquote {\bibinfo {title} {Stochastic
  basins of attraction and generalized committor functions},}\ }\href@noop {}
  {\bibfield  {journal} {\bibinfo  {journal} {Physical Review E}\ }\textbf
  {\bibinfo {volume} {100}},\ \bibinfo {pages} {022124} (\bibinfo {year}
  {2019})}\BibitemShut {NoStop}%
\bibitem [{\citenamefont {Kim}, \citenamefont {Lee},\ and\ \citenamefont
  {Holme}(2016)}]{kim2016building}%
  \BibitemOpen
  \bibfield  {author} {\bibinfo {author} {\bibfnamefont {H.}~\bibnamefont
  {Kim}}, \bibinfo {author} {\bibfnamefont {S.~H.}\ \bibnamefont {Lee}}, \ and\
  \bibinfo {author} {\bibfnamefont {P.}~\bibnamefont {Holme}},\ }\bibfield
  {title} {\enquote {\bibinfo {title} {Building blocks of the basin stability
  of power grids},}\ }\href@noop {} {\bibfield  {journal} {\bibinfo  {journal}
  {Physical Review E}\ }\textbf {\bibinfo {volume} {93}},\ \bibinfo {pages}
  {062318} (\bibinfo {year} {2016})}\BibitemShut {NoStop}%
\bibitem [{\citenamefont {Kim}\ \emph {et~al.}(2018)\citenamefont {Kim},
  \citenamefont {Lee}, \citenamefont {Davidsen},\ and\ \citenamefont
  {Son}}]{kim2018multistability}%
  \BibitemOpen
  \bibfield  {author} {\bibinfo {author} {\bibfnamefont {H.}~\bibnamefont
  {Kim}}, \bibinfo {author} {\bibfnamefont {S.~H.}\ \bibnamefont {Lee}},
  \bibinfo {author} {\bibfnamefont {J.}~\bibnamefont {Davidsen}}, \ and\
  \bibinfo {author} {\bibfnamefont {S.-W.}\ \bibnamefont {Son}},\ }\bibfield
  {title} {\enquote {\bibinfo {title} {Multistability and variations in basin
  of attraction in power-grid systems},}\ }\href@noop {} {\bibfield  {journal}
  {\bibinfo  {journal} {New Journal of Physics}\ }\textbf {\bibinfo {volume}
  {20}},\ \bibinfo {pages} {113006} (\bibinfo {year} {2018})}\BibitemShut
  {NoStop}%
\bibitem [{\citenamefont {Wolff}, \citenamefont {Lind},\ and\ \citenamefont
  {Maass}(2018)}]{wolff2018power}%
  \BibitemOpen
  \bibfield  {author} {\bibinfo {author} {\bibfnamefont {M.~F.}\ \bibnamefont
  {Wolff}}, \bibinfo {author} {\bibfnamefont {P.~G.}\ \bibnamefont {Lind}}, \
  and\ \bibinfo {author} {\bibfnamefont {P.}~\bibnamefont {Maass}},\ }\bibfield
   {title} {\enquote {\bibinfo {title} {Power grid stability under perturbation
  of single nodes: Effects of heterogeneity and internal nodes},}\ }\href@noop
  {} {\bibfield  {journal} {\bibinfo  {journal} {Chaos: An Interdisciplinary
  Journal of Nonlinear Science}\ }\textbf {\bibinfo {volume} {28}},\ \bibinfo
  {pages} {103120} (\bibinfo {year} {2018})}\BibitemShut {NoStop}%
\bibitem [{\citenamefont {Donges}\ \emph {et~al.}(2011)\citenamefont {Donges},
  \citenamefont {Schultz}, \citenamefont {Marwan}, \citenamefont {Zou},\ and\
  \citenamefont {Kurths}}]{Donges2011}%
  \BibitemOpen
  \bibfield  {author} {\bibinfo {author} {\bibfnamefont {J.~F.}\ \bibnamefont
  {Donges}}, \bibinfo {author} {\bibfnamefont {H.~C.}\ \bibnamefont {Schultz}},
  \bibinfo {author} {\bibfnamefont {N.}~\bibnamefont {Marwan}}, \bibinfo
  {author} {\bibfnamefont {Y.}~\bibnamefont {Zou}}, \ and\ \bibinfo {author}
  {\bibfnamefont {J.}~\bibnamefont {Kurths}},\ }\bibfield  {title} {\enquote
  {\bibinfo {title} {{Investigating the topology of interacting networks:
  Theory and application to coupled climate subnetworks}},}\ }\href@noop {}
  {\bibfield  {journal} {\bibinfo  {journal} {European Physical Journal B}\
  }\textbf {\bibinfo {volume} {84}},\ \bibinfo {pages} {635--651} (\bibinfo
  {year} {2011})}\BibitemShut {NoStop}%
\bibitem [{\citenamefont {Criado}\ \emph {et~al.}(2012)\citenamefont {Criado},
  \citenamefont {Flores}, \citenamefont {{Garc{\'{i}}a Del Amo}}, \citenamefont
  {G{\'{o}}mez-Garde{\~{n}}es},\ and\ \citenamefont {Romance}}]{Criado2012}%
  \BibitemOpen
  \bibfield  {author} {\bibinfo {author} {\bibfnamefont {R.}~\bibnamefont
  {Criado}}, \bibinfo {author} {\bibfnamefont {J.}~\bibnamefont {Flores}},
  \bibinfo {author} {\bibfnamefont {A.}~\bibnamefont {{Garc{\'{i}}a Del Amo}}},
  \bibinfo {author} {\bibfnamefont {J.}~\bibnamefont
  {G{\'{o}}mez-Garde{\~{n}}es}}, \ and\ \bibinfo {author} {\bibfnamefont
  {M.}~\bibnamefont {Romance}},\ }\bibfield  {title} {\enquote {\bibinfo
  {title} {{A mathematical model for networks with structures in the
  mesoscale}},}\ }\href@noop {} {\bibfield  {journal} {\bibinfo  {journal}
  {International Journal of Computer Mathematics}\ }\textbf {\bibinfo {volume}
  {89}},\ \bibinfo {pages} {291--309} (\bibinfo {year} {2012})}\BibitemShut
  {NoStop}%
\bibitem [{\citenamefont {Schneider}\ \emph {et~al.}(2013)\citenamefont
  {Schneider}, \citenamefont {Yazdani}, \citenamefont {Ara{\'{u}}jo},
  \citenamefont {Havlin},\ and\ \citenamefont {Herrmann}}]{Schneider2011}%
  \BibitemOpen
  \bibfield  {author} {\bibinfo {author} {\bibfnamefont {C.~M.}\ \bibnamefont
  {Schneider}}, \bibinfo {author} {\bibfnamefont {N.}~\bibnamefont {Yazdani}},
  \bibinfo {author} {\bibfnamefont {N.~A.}\ \bibnamefont {Ara{\'{u}}jo}},
  \bibinfo {author} {\bibfnamefont {S.}~\bibnamefont {Havlin}}, \ and\ \bibinfo
  {author} {\bibfnamefont {H.~J.}\ \bibnamefont {Herrmann}},\ }\bibfield
  {title} {\enquote {\bibinfo {title} {{Towards designing robust coupled
  networks}},}\ }\href@noop {} {\bibfield  {journal} {\bibinfo  {journal}
  {Scientific Reports}\ }\textbf {\bibinfo {volume} {3}},\ \bibinfo {pages}
  {1969} (\bibinfo {year} {2013})}\BibitemShut {NoStop}%
\bibitem [{\citenamefont {Buldyrev}\ \emph {et~al.}(2010)\citenamefont
  {Buldyrev}, \citenamefont {Parshani}, \citenamefont {Paul}, \citenamefont
  {Stanley},\ and\ \citenamefont {Havlin}}]{Buldyrev2010b}%
  \BibitemOpen
  \bibfield  {author} {\bibinfo {author} {\bibfnamefont {S.~V.}\ \bibnamefont
  {Buldyrev}}, \bibinfo {author} {\bibfnamefont {R.}~\bibnamefont {Parshani}},
  \bibinfo {author} {\bibfnamefont {G.}~\bibnamefont {Paul}}, \bibinfo {author}
  {\bibfnamefont {H.~E.}\ \bibnamefont {Stanley}}, \ and\ \bibinfo {author}
  {\bibfnamefont {S.}~\bibnamefont {Havlin}},\ }\bibfield  {title} {\enquote
  {\bibinfo {title} {{Catastrophic cascade of failures in interdependent
  networks}},}\ }\href@noop {} {\bibfield  {journal} {\bibinfo  {journal}
  {Nature}\ }\textbf {\bibinfo {volume} {464}},\ \bibinfo {pages} {1025--1028}
  (\bibinfo {year} {2010})}\BibitemShut {NoStop}%
\bibitem [{\citenamefont {Vespignani}(2010)}]{Vespignani2010}%
  \BibitemOpen
  \bibfield  {author} {\bibinfo {author} {\bibfnamefont {A.}~\bibnamefont
  {Vespignani}},\ }\bibfield  {title} {\enquote {\bibinfo {title} {{Complex
  networks: The fragility of interdependency}},}\ }\href@noop {} {\bibfield
  {journal} {\bibinfo  {journal} {Nature}\ }\textbf {\bibinfo {volume} {464}},\
  \bibinfo {pages} {984--985} (\bibinfo {year} {2010})}\BibitemShut {NoStop}%
\bibitem [{\citenamefont {Krishna}\ \emph {et~al.}(2019)\citenamefont
  {Krishna}, \citenamefont {Schiffer}, \citenamefont {Monshizadeh},\ and\
  \citenamefont {Raisch}}]{krishna2019consensus}%
  \BibitemOpen
  \bibfield  {author} {\bibinfo {author} {\bibfnamefont {A.}~\bibnamefont
  {Krishna}}, \bibinfo {author} {\bibfnamefont {J.}~\bibnamefont {Schiffer}},
  \bibinfo {author} {\bibfnamefont {N.}~\bibnamefont {Monshizadeh}}, \ and\
  \bibinfo {author} {\bibfnamefont {J.}~\bibnamefont {Raisch}},\ }\bibfield
  {title} {\enquote {\bibinfo {title} {A consensus-based voltage control for
  reactive power sharing and {PC}c voltage regulation in microgrids with
  parallel-connected inverters},}\ }in\ \href@noop {} {\emph {\bibinfo
  {booktitle} {2019 18\textsuperscript{th} European Control Conference
  (ECC)}}}\ (\bibinfo {organization} {IEEE},\ \bibinfo {year} {2019})\ pp.\
  \bibinfo {pages} {542--547}\BibitemShut {NoStop}%
\bibitem [{\citenamefont {Mahmoud}, \citenamefont {Rahman},\ and\ \citenamefont
  {Fouad}(2015)}]{mahmoud2015review}%
  \BibitemOpen
  \bibfield  {author} {\bibinfo {author} {\bibfnamefont {M.~S.}\ \bibnamefont
  {Mahmoud}}, \bibinfo {author} {\bibfnamefont {M.~S.~U.}\ \bibnamefont
  {Rahman}}, \ and\ \bibinfo {author} {\bibfnamefont {M.-S.}\ \bibnamefont
  {Fouad}},\ }\bibfield  {title} {\enquote {\bibinfo {title} {Review of
  microgrid architectures--a system of systems perspective},}\ }\href@noop {}
  {\bibfield  {journal} {\bibinfo  {journal} {IET Renewable Power Generation}\
  }\textbf {\bibinfo {volume} {9}},\ \bibinfo {pages} {1064--1078} (\bibinfo
  {year} {2015})}\BibitemShut {NoStop}%
\bibitem [{\citenamefont {Wu}\ \emph {et~al.}(2014)\citenamefont {Wu},
  \citenamefont {Dragicevic}, \citenamefont {Vasquez}, \citenamefont
  {Guerrero},\ and\ \citenamefont {Guan}}]{wu2014secondary}%
  \BibitemOpen
  \bibfield  {author} {\bibinfo {author} {\bibfnamefont {D.}~\bibnamefont
  {Wu}}, \bibinfo {author} {\bibfnamefont {T.}~\bibnamefont {Dragicevic}},
  \bibinfo {author} {\bibfnamefont {J.~C.}\ \bibnamefont {Vasquez}}, \bibinfo
  {author} {\bibfnamefont {J.~M.}\ \bibnamefont {Guerrero}}, \ and\ \bibinfo
  {author} {\bibfnamefont {Y.}~\bibnamefont {Guan}},\ }\bibfield  {title}
  {\enquote {\bibinfo {title} {Secondary coordinated control of islanded
  microgrids based on consensus algorithms},}\ }in\ \href@noop {} {\emph
  {\bibinfo {booktitle} {2014 IEEE Energy Conversion Congress and Exposition
  (ECCE)}}}\ (\bibinfo {organization} {IEEE},\ \bibinfo {year} {2014})\ pp.\
  \bibinfo {pages} {4290--4297}\BibitemShut {NoStop}%
\bibitem [{\citenamefont {Mao}\ \emph {et~al.}(2019)\citenamefont {Mao},
  \citenamefont {Dong}, \citenamefont {Schultz}, \citenamefont {Tang},
  \citenamefont {Meng}, \citenamefont {Dong},\ and\ \citenamefont
  {Qian}}]{Mao2019}%
  \BibitemOpen
  \bibfield  {author} {\bibinfo {author} {\bibfnamefont {S.}~\bibnamefont
  {Mao}}, \bibinfo {author} {\bibfnamefont {Z.}~\bibnamefont {Dong}}, \bibinfo
  {author} {\bibfnamefont {P.}~\bibnamefont {Schultz}}, \bibinfo {author}
  {\bibfnamefont {Y.}~\bibnamefont {Tang}}, \bibinfo {author} {\bibfnamefont
  {K.}~\bibnamefont {Meng}}, \bibinfo {author} {\bibfnamefont {Z.~Y.}\
  \bibnamefont {Dong}}, \ and\ \bibinfo {author} {\bibfnamefont
  {F.}~\bibnamefont {Qian}},\ }\bibfield  {title} {\enquote {\bibinfo {title}
  {A finite-time distributed optimization algorithm for economic dispatch in
  smart grids},}\ }\href@noop {} {\bibfield  {journal} {\bibinfo  {journal}
  {IEEE Transactions on Systems, Man, and Cybernetics: Systems}\ } (\bibinfo
  {year} {2019})}\BibitemShut {NoStop}%
\bibitem [{\citenamefont {{Weitenberg}}\ \emph {et~al.}(2018)\citenamefont
  {{Weitenberg}}, \citenamefont {{Jiang}}, \citenamefont {{Zhao}},
  \citenamefont {{Mallada}}, \citenamefont {{D\"orfler}},\ and\ \citenamefont
  {{De Persis}}}]{doerfler2018leakyintegrator}%
  \BibitemOpen
  \bibfield  {author} {\bibinfo {author} {\bibfnamefont {E.}~\bibnamefont
  {{Weitenberg}}}, \bibinfo {author} {\bibfnamefont {Y.}~\bibnamefont
  {{Jiang}}}, \bibinfo {author} {\bibfnamefont {C.}~\bibnamefont {{Zhao}}},
  \bibinfo {author} {\bibfnamefont {E.}~\bibnamefont {{Mallada}}}, \bibinfo
  {author} {\bibfnamefont {F.}~\bibnamefont {{D\"orfler}}}, \ and\ \bibinfo
  {author} {\bibfnamefont {C.}~\bibnamefont {{De Persis}}},\ }\bibfield
  {title} {\enquote {\bibinfo {title} {Robust decentralized frequency control:
  A leaky integrator approach},}\ }in\ \href@noop {} {\emph {\bibinfo
  {booktitle} {2018 European Control Conference (ECC)}}}\ (\bibinfo {year}
  {2018})\ pp.\ \bibinfo {pages} {764--769}\BibitemShut {NoStop}%
\bibitem [{\citenamefont {Weidlich}\ and\ \citenamefont
  {Veit}(2008)}]{weidlich2008critical}%
  \BibitemOpen
  \bibfield  {author} {\bibinfo {author} {\bibfnamefont {A.}~\bibnamefont
  {Weidlich}}\ and\ \bibinfo {author} {\bibfnamefont {D.}~\bibnamefont
  {Veit}},\ }\bibfield  {title} {\enquote {\bibinfo {title} {A critical survey
  of agent-based wholesale electricity market models},}\ }\href@noop {}
  {\bibfield  {journal} {\bibinfo  {journal} {Energy Economics}\ }\textbf
  {\bibinfo {volume} {30}},\ \bibinfo {pages} {1728--1759} (\bibinfo {year}
  {2008})}\BibitemShut {NoStop}%
\bibitem [{\citenamefont {Ringler}, \citenamefont {Keles},\ and\ \citenamefont
  {Fichtner}(2016)}]{ringler2016agent}%
  \BibitemOpen
  \bibfield  {author} {\bibinfo {author} {\bibfnamefont {P.}~\bibnamefont
  {Ringler}}, \bibinfo {author} {\bibfnamefont {D.}~\bibnamefont {Keles}}, \
  and\ \bibinfo {author} {\bibfnamefont {W.}~\bibnamefont {Fichtner}},\
  }\bibfield  {title} {\enquote {\bibinfo {title} {Agent-based modelling and
  simulation of smart electricity grids and markets--a literature review},}\
  }\href@noop {} {\bibfield  {journal} {\bibinfo  {journal} {Renewable and
  Sustainable Energy Reviews}\ }\textbf {\bibinfo {volume} {57}},\ \bibinfo
  {pages} {205--215} (\bibinfo {year} {2016})}\BibitemShut {NoStop}%
\bibitem [{\citenamefont {Mureddu}\ and\ \citenamefont
  {Meyer-Ortmanns}(2018)}]{mureddu2018extreme}%
  \BibitemOpen
  \bibfield  {author} {\bibinfo {author} {\bibfnamefont {M.}~\bibnamefont
  {Mureddu}}\ and\ \bibinfo {author} {\bibfnamefont {H.}~\bibnamefont
  {Meyer-Ortmanns}},\ }\bibfield  {title} {\enquote {\bibinfo {title} {Extreme
  prices in electricity balancing markets from an approach of statistical
  physics},}\ }\href@noop {} {\bibfield  {journal} {\bibinfo  {journal}
  {Physica A: Statistical Mechanics and its Applications}\ }\textbf {\bibinfo
  {volume} {490}},\ \bibinfo {pages} {1324--1334} (\bibinfo {year}
  {2018})}\BibitemShut {NoStop}%
\bibitem [{\citenamefont {Claessen}\ \emph {et~al.}(2014)\citenamefont
  {Claessen}, \citenamefont {Claessens}, \citenamefont {Hommelberg},
  \citenamefont {Molderink}, \citenamefont {Bakker}, \citenamefont {Toersche},\
  and\ \citenamefont {van~den Broek}}]{claessen2014comparative}%
  \BibitemOpen
  \bibfield  {author} {\bibinfo {author} {\bibfnamefont {F.}~\bibnamefont
  {Claessen}}, \bibinfo {author} {\bibfnamefont {B.}~\bibnamefont {Claessens}},
  \bibinfo {author} {\bibfnamefont {M.}~\bibnamefont {Hommelberg}}, \bibinfo
  {author} {\bibfnamefont {A.}~\bibnamefont {Molderink}}, \bibinfo {author}
  {\bibfnamefont {V.}~\bibnamefont {Bakker}}, \bibinfo {author} {\bibfnamefont
  {H.}~\bibnamefont {Toersche}}, \ and\ \bibinfo {author} {\bibfnamefont
  {M.}~\bibnamefont {van~den Broek}},\ }\bibfield  {title} {\enquote {\bibinfo
  {title} {Comparative analysis of tertiary control systems for smart grids
  using the flex street model},}\ }\href@noop {} {\bibfield  {journal}
  {\bibinfo  {journal} {Renewable Energy}\ }\textbf {\bibinfo {volume} {69}},\
  \bibinfo {pages} {260--270} (\bibinfo {year} {2014})}\BibitemShut {NoStop}%
\bibitem [{\citenamefont {Zhao}\ \emph {et~al.}(2016)\citenamefont {Zhao},
  \citenamefont {Mallada}, \citenamefont {Low},\ and\ \citenamefont
  {Bialek}}]{zhao2016unified}%
  \BibitemOpen
  \bibfield  {author} {\bibinfo {author} {\bibfnamefont {C.}~\bibnamefont
  {Zhao}}, \bibinfo {author} {\bibfnamefont {E.}~\bibnamefont {Mallada}},
  \bibinfo {author} {\bibfnamefont {S.}~\bibnamefont {Low}}, \ and\ \bibinfo
  {author} {\bibfnamefont {J.}~\bibnamefont {Bialek}},\ }\bibfield  {title}
  {\enquote {\bibinfo {title} {A unified framework for frequency control and
  congestion management},}\ }in\ \href@noop {} {\emph {\bibinfo {booktitle}
  {2016 Power Systems Computation Conference (PSCC)}}}\ (\bibinfo
  {organization} {IEEE},\ \bibinfo {year} {2016})\ pp.\ \bibinfo {pages}
  {1--7}\BibitemShut {NoStop}%
\bibitem [{\citenamefont {Amann}, \citenamefont {Owens},\ and\ \citenamefont
  {Rogers}(1998)}]{amann1998predictive}%
  \BibitemOpen
  \bibfield  {author} {\bibinfo {author} {\bibfnamefont {N.}~\bibnamefont
  {Amann}}, \bibinfo {author} {\bibfnamefont {D.~H.}\ \bibnamefont {Owens}}, \
  and\ \bibinfo {author} {\bibfnamefont {E.}~\bibnamefont {Rogers}},\
  }\bibfield  {title} {\enquote {\bibinfo {title} {Predictive optimal iterative
  learning control},}\ }\href@noop {} {\bibfield  {journal} {\bibinfo
  {journal} {International Journal of Control}\ }\textbf {\bibinfo {volume}
  {69}},\ \bibinfo {pages} {203--226} (\bibinfo {year} {1998})}\BibitemShut
  {NoStop}%
\bibitem [{\citenamefont {Bristow}, \citenamefont {Tharayil},\ and\
  \citenamefont {Alleyne}(2006)}]{bristow2006survey}%
  \BibitemOpen
  \bibfield  {author} {\bibinfo {author} {\bibfnamefont {D.~A.}\ \bibnamefont
  {Bristow}}, \bibinfo {author} {\bibfnamefont {M.}~\bibnamefont {Tharayil}}, \
  and\ \bibinfo {author} {\bibfnamefont {A.~G.}\ \bibnamefont {Alleyne}},\
  }\bibfield  {title} {\enquote {\bibinfo {title} {A survey of iterative
  learning control},}\ }\href@noop {} {\bibfield  {journal} {\bibinfo
  {journal} {IEEE Control Systems Magazine}\ }\textbf {\bibinfo {volume}
  {26}},\ \bibinfo {pages} {96--114} (\bibinfo {year} {2006})}\BibitemShut
  {NoStop}%
\bibitem [{\citenamefont {Zeng}\ \emph {et~al.}(2013)\citenamefont {Zeng},
  \citenamefont {Yang}, \citenamefont {Zhao},\ and\ \citenamefont
  {Cheng}}]{zeng2013topologies}%
  \BibitemOpen
  \bibfield  {author} {\bibinfo {author} {\bibfnamefont {Z.}~\bibnamefont
  {Zeng}}, \bibinfo {author} {\bibfnamefont {H.}~\bibnamefont {Yang}}, \bibinfo
  {author} {\bibfnamefont {R.}~\bibnamefont {Zhao}}, \ and\ \bibinfo {author}
  {\bibfnamefont {C.}~\bibnamefont {Cheng}},\ }\bibfield  {title} {\enquote
  {\bibinfo {title} {Topologies and control strategies of multi-functional
  grid-connected inverters for power quality enhancement: A comprehensive
  review},}\ }\href@noop {} {\bibfield  {journal} {\bibinfo  {journal}
  {Renewable and Sustainable Energy Reviews}\ }\textbf {\bibinfo {volume}
  {24}},\ \bibinfo {pages} {223--270} (\bibinfo {year} {2013})}\BibitemShut
  {NoStop}%
\bibitem [{\citenamefont {Teng}(2014)}]{teng2014repetitive}%
  \BibitemOpen
  \bibfield  {author} {\bibinfo {author} {\bibfnamefont {K.-T.}\ \bibnamefont
  {Teng}},\ }\emph {\bibinfo {title} {Repetitive and Iterative Learning Control
  for Power Converter and Precision Motion Control}},\ \href@noop {} {Ph.D.
  thesis},\ \bibinfo  {school} {UCLA} (\bibinfo {year} {2014})\BibitemShut
  {NoStop}%
\bibitem [{\citenamefont {Chai}\ \emph {et~al.}(2016)\citenamefont {Chai},
  \citenamefont {Yang}, \citenamefont {Gao},\ and\ \citenamefont
  {Zhao}}]{chai2016}%
  \BibitemOpen
  \bibfield  {author} {\bibinfo {author} {\bibfnamefont {B.}~\bibnamefont
  {Chai}}, \bibinfo {author} {\bibfnamefont {Z.}~\bibnamefont {Yang}}, \bibinfo
  {author} {\bibfnamefont {K.}~\bibnamefont {Gao}}, \ and\ \bibinfo {author}
  {\bibfnamefont {T.}~\bibnamefont {Zhao}},\ }\bibfield  {title} {\enquote
  {\bibinfo {title} {Iterative learning for optimal residential load scheduling
  in smart grid},}\ }\href@noop {} {\bibfield  {journal} {\bibinfo  {journal}
  {Ad Hoc Networks}\ }\textbf {\bibinfo {volume} {41}},\ \bibinfo {pages}
  {99--111} (\bibinfo {year} {2016})}\BibitemShut {NoStop}%
\bibitem [{\citenamefont {Bampoulas}\ \emph {et~al.}(2019)\citenamefont
  {Bampoulas}, \citenamefont {Saffari}, \citenamefont {Pallonetto},
  \citenamefont {Mangina},\ and\ \citenamefont {Finn}}]{bampoulas2019self}%
  \BibitemOpen
  \bibfield  {author} {\bibinfo {author} {\bibfnamefont {A.}~\bibnamefont
  {Bampoulas}}, \bibinfo {author} {\bibfnamefont {M.}~\bibnamefont {Saffari}},
  \bibinfo {author} {\bibfnamefont {F.}~\bibnamefont {Pallonetto}}, \bibinfo
  {author} {\bibfnamefont {E.}~\bibnamefont {Mangina}}, \ and\ \bibinfo
  {author} {\bibfnamefont {D.~P.}\ \bibnamefont {Finn}},\ }\bibfield  {title}
  {\enquote {\bibinfo {title} {Self-learning control algorithms for energy
  systems integration in the residential building sector},}\ }in\ \href@noop {}
  {\emph {\bibinfo {booktitle} {2019 IEEE 5\textsuperscript{th} World Forum on
  Internet of Things (WF-IoT)}}}\ (\bibinfo {organization} {IEEE},\ \bibinfo
  {year} {2019})\ pp.\ \bibinfo {pages} {815--818}\BibitemShut {NoStop}%
\bibitem [{\citenamefont {Guo}\ \emph {et~al.}(2015)\citenamefont {Guo},
  \citenamefont {Liu}, \citenamefont {Si}, \citenamefont {He}, \citenamefont
  {Harley},\ and\ \citenamefont {Mei}}]{Guo2016}%
  \BibitemOpen
  \bibfield  {author} {\bibinfo {author} {\bibfnamefont {W.}~\bibnamefont
  {Guo}}, \bibinfo {author} {\bibfnamefont {F.}~\bibnamefont {Liu}}, \bibinfo
  {author} {\bibfnamefont {J.}~\bibnamefont {Si}}, \bibinfo {author}
  {\bibfnamefont {D.}~\bibnamefont {He}}, \bibinfo {author} {\bibfnamefont
  {R.}~\bibnamefont {Harley}}, \ and\ \bibinfo {author} {\bibfnamefont
  {S.}~\bibnamefont {Mei}},\ }\bibfield  {title} {\enquote {\bibinfo {title}
  {Online supplementary {ADP} learning controller design and application to
  power system frequency control with large-scale wind energy integration},}\
  }\href@noop {} {\bibfield  {journal} {\bibinfo  {journal} {IEEE Transactions
  on Neural Networks and Learning Systems}\ }\textbf {\bibinfo {volume} {27}},\
  \bibinfo {pages} {1748--1761} (\bibinfo {year} {2015})}\BibitemShut {NoStop}%
\bibitem [{\citenamefont {Guo}\ \emph {et~al.}(2019)\citenamefont {Guo},
  \citenamefont {Liu}, \citenamefont {Yong}, \citenamefont {Cheng},\ and\
  \citenamefont {Muhammad}}]{Guo2019}%
  \BibitemOpen
  \bibfield  {author} {\bibinfo {author} {\bibfnamefont {H.-Q.}\ \bibnamefont
  {Guo}}, \bibinfo {author} {\bibfnamefont {C.-Z.}\ \bibnamefont {Liu}},
  \bibinfo {author} {\bibfnamefont {J.-W.}\ \bibnamefont {Yong}}, \bibinfo
  {author} {\bibfnamefont {X.-Q.}\ \bibnamefont {Cheng}}, \ and\ \bibinfo
  {author} {\bibfnamefont {F.}~\bibnamefont {Muhammad}},\ }\bibfield  {title}
  {\enquote {\bibinfo {title} {Model predictive iterative learning control for
  energy management of plug-in hybrid electric vehicle},}\ }\href@noop {}
  {\bibfield  {journal} {\bibinfo  {journal} {IEEE Access}\ } (\bibinfo {year}
  {2019})}\BibitemShut {NoStop}%
\bibitem [{\citenamefont {Nguyen}\ and\ \citenamefont
  {Banjerdpongchai}(2016)}]{nguyen2016iterative}%
  \BibitemOpen
  \bibfield  {author} {\bibinfo {author} {\bibfnamefont {D.~H.}\ \bibnamefont
  {Nguyen}}\ and\ \bibinfo {author} {\bibfnamefont {D.}~\bibnamefont
  {Banjerdpongchai}},\ }\bibfield  {title} {\enquote {\bibinfo {title}
  {Iterative learning control of energy management system: Survey on
  multi-agent system framework},}\ }\href@noop {} {\bibfield  {journal}
  {\bibinfo  {journal} {Engineering Journal}\ }\textbf {\bibinfo {volume}
  {20}},\ \bibinfo {pages} {1--4} (\bibinfo {year} {2016})}\BibitemShut
  {NoStop}%
\bibitem [{\citenamefont {Xu}\ and\ \citenamefont
  {Yang}(2013)}]{xu2013iterative}%
  \BibitemOpen
  \bibfield  {author} {\bibinfo {author} {\bibfnamefont {J.-X.}\ \bibnamefont
  {Xu}}\ and\ \bibinfo {author} {\bibfnamefont {S.}~\bibnamefont {Yang}},\
  }\bibfield  {title} {\enquote {\bibinfo {title} {Iterative learning based
  control and optimization for large scale systems},}\ }\href@noop {}
  {\bibfield  {journal} {\bibinfo  {journal} {IFAC Proceedings Volumes}\
  }\textbf {\bibinfo {volume} {46}},\ \bibinfo {pages} {74--81} (\bibinfo
  {year} {2013})}\BibitemShut {NoStop}%
\bibitem [{\citenamefont {Strenge}\ \emph {et~al.}(2020)\citenamefont
  {Strenge}, \citenamefont {Jing}, \citenamefont {Seel}, \citenamefont
  {Hellmann},\ and\ \citenamefont {Raisch}}]{Strenge2019}%
  \BibitemOpen
  \bibfield  {author} {\bibinfo {author} {\bibfnamefont {L.}~\bibnamefont
  {Strenge}}, \bibinfo {author} {\bibfnamefont {X.}~\bibnamefont {Jing}},
  \bibinfo {author} {\bibfnamefont {T.}~\bibnamefont {Seel}}, \bibinfo {author}
  {\bibfnamefont {F.}~\bibnamefont {Hellmann}}, \ and\ \bibinfo {author}
  {\bibfnamefont {J.}~\bibnamefont {Raisch}},\ }\bibfield  {title} {\enquote
  {\bibinfo {title} {Iterative learning control in prosumer-based microgrids
  with hierarchical control},}\ }in\ \href@noop {} {\emph {\bibinfo {booktitle}
  {submitted to IFAC World Congress 2020}}}\ (\bibinfo {year} {2020})\ p.\
  \bibinfo {pages} {tbd}\BibitemShut {NoStop}%
\bibitem [{\citenamefont {Hill}\ and\ \citenamefont {Chen}(2006)}]{Hill2006}%
  \BibitemOpen
  \bibfield  {author} {\bibinfo {author} {\bibfnamefont {D.~J.}\ \bibnamefont
  {Hill}}\ and\ \bibinfo {author} {\bibfnamefont {G.}~\bibnamefont {Chen}},\
  }\bibfield  {title} {\enquote {\bibinfo {title} {Power systems as dynamic
  networks},}\ }in\ \href@noop {} {\emph {\bibinfo {booktitle} {2006 IEEE
  International Symposium on Circuits and Systems}}}\ (\bibinfo {organization}
  {IEEE},\ \bibinfo {year} {2006})\ pp.\ \bibinfo {pages} {4--pp}\BibitemShut
  {NoStop}%
\bibitem [{\citenamefont {Machowski}, \citenamefont {Bialek},\ and\
  \citenamefont {Bumby}(2011)}]{Machowski2011}%
  \BibitemOpen
  \bibfield  {author} {\bibinfo {author} {\bibfnamefont {J.}~\bibnamefont
  {Machowski}}, \bibinfo {author} {\bibfnamefont {J.}~\bibnamefont {Bialek}}, \
  and\ \bibinfo {author} {\bibfnamefont {J.}~\bibnamefont {Bumby}},\
  }\href@noop {} {\emph {\bibinfo {title} {{Power system dynamics: stability
  and control}}}}\ (\bibinfo  {publisher} {John Wiley $\backslash${\&} Sons,
  Ltd.},\ \bibinfo {year} {2011})\BibitemShut {NoStop}%
\bibitem [{\citenamefont {Schiffer}\ \emph {et~al.}(2014)\citenamefont
  {Schiffer}, \citenamefont {Ortega}, \citenamefont {Astolfi}, \citenamefont
  {Raisch},\ and\ \citenamefont {Sezi}}]{schiffer-cond}%
  \BibitemOpen
  \bibfield  {author} {\bibinfo {author} {\bibfnamefont {J.}~\bibnamefont
  {Schiffer}}, \bibinfo {author} {\bibfnamefont {R.}~\bibnamefont {Ortega}},
  \bibinfo {author} {\bibfnamefont {A.}~\bibnamefont {Astolfi}}, \bibinfo
  {author} {\bibfnamefont {J.}~\bibnamefont {Raisch}}, \ and\ \bibinfo {author}
  {\bibfnamefont {T.}~\bibnamefont {Sezi}},\ }\bibfield  {title} {\enquote
  {\bibinfo {title} {Conditions for stability of droop-controlled
  inverter-based microgrids},}\ }\href@noop {} {\bibfield  {journal} {\bibinfo
  {journal} {Automatica}\ }\textbf {\bibinfo {volume} {50}},\ \bibinfo {pages}
  {2457--2469} (\bibinfo {year} {2014})}\BibitemShut {NoStop}%
\bibitem [{\citenamefont {Sch{\"a}fer}\ \emph {et~al.}(2018)\citenamefont
  {Sch{\"a}fer}, \citenamefont {Beck}, \citenamefont {Aihara}, \citenamefont
  {Witthaut},\ and\ \citenamefont {Timme}}]{schafer2018non}%
  \BibitemOpen
  \bibfield  {author} {\bibinfo {author} {\bibfnamefont {B.}~\bibnamefont
  {Sch{\"a}fer}}, \bibinfo {author} {\bibfnamefont {C.}~\bibnamefont {Beck}},
  \bibinfo {author} {\bibfnamefont {K.}~\bibnamefont {Aihara}}, \bibinfo
  {author} {\bibfnamefont {D.}~\bibnamefont {Witthaut}}, \ and\ \bibinfo
  {author} {\bibfnamefont {M.}~\bibnamefont {Timme}},\ }\bibfield  {title}
  {\enquote {\bibinfo {title} {Non-{Gauss}ian power grid frequency fluctuations
  characterized by l{\'e}vy-stable laws and superstatistics},}\ }\href@noop {}
  {\bibfield  {journal} {\bibinfo  {journal} {Nature Energy}\ }\textbf
  {\bibinfo {volume} {3}},\ \bibinfo {pages} {119} (\bibinfo {year}
  {2018})}\BibitemShut {NoStop}%
\bibitem [{\citenamefont {Anvari}\ \emph {et~al.}(2019)\citenamefont {Anvari},
  \citenamefont {Gorj{\~a}o}, \citenamefont {Timme}, \citenamefont {Witthaut},
  \citenamefont {Sch{\"a}fer},\ and\ \citenamefont
  {Kantz}}]{anvari2019stochastic}%
  \BibitemOpen
  \bibfield  {author} {\bibinfo {author} {\bibfnamefont {M.}~\bibnamefont
  {Anvari}}, \bibinfo {author} {\bibfnamefont {L.~R.}\ \bibnamefont
  {Gorj{\~a}o}}, \bibinfo {author} {\bibfnamefont {M.}~\bibnamefont {Timme}},
  \bibinfo {author} {\bibfnamefont {D.}~\bibnamefont {Witthaut}}, \bibinfo
  {author} {\bibfnamefont {B.}~\bibnamefont {Sch{\"a}fer}}, \ and\ \bibinfo
  {author} {\bibfnamefont {H.}~\bibnamefont {Kantz}},\ }\bibfield  {title}
  {\enquote {\bibinfo {title} {Stochastic properties of the frequency dynamics
  in real and synthetic power grids},}\ }\href@noop {} {\bibfield  {journal}
  {\bibinfo  {journal} {arXiv:1909.09110}\ } (\bibinfo {year}
  {2019})}\BibitemShut {NoStop}%
\bibitem [{\citenamefont {Nicosia}\ \emph {et~al.}(2017)\citenamefont
  {Nicosia}, \citenamefont {Skardal}, \citenamefont {Arenas},\ and\
  \citenamefont {Latora}}]{Nicosia2014a}%
  \BibitemOpen
  \bibfield  {author} {\bibinfo {author} {\bibfnamefont {V.}~\bibnamefont
  {Nicosia}}, \bibinfo {author} {\bibfnamefont {P.~S.}\ \bibnamefont
  {Skardal}}, \bibinfo {author} {\bibfnamefont {A.}~\bibnamefont {Arenas}}, \
  and\ \bibinfo {author} {\bibfnamefont {V.}~\bibnamefont {Latora}},\
  }\bibfield  {title} {\enquote {\bibinfo {title} {{Collective Phenomena
  Emerging from the Interactions between Dynamical Processes in Multiplex
  Networks}},}\ }\href@noop {} {\bibfield  {journal} {\bibinfo  {journal}
  {Physical Review Letters}\ }\textbf {\bibinfo {volume} {118}},\ \bibinfo
  {pages} {138302} (\bibinfo {year} {2017})}\BibitemShut {NoStop}%
\bibitem [{Note1()}]{Note1}%
  \BibitemOpen
  \bibinfo {note} {Repository controlled-multi-timescale-powergrid at \protect
  \url
  {https://github.com/strangeli/controlled-multi-timescale-powergrid}}\BibitemShut
  {NoStop}%
\bibitem [{\citenamefont {Rackauckas}\ and\ \citenamefont
  {Nie}(2017)}]{rackauckas2017differentialequations}%
  \BibitemOpen
  \bibfield  {author} {\bibinfo {author} {\bibfnamefont {C.}~\bibnamefont
  {Rackauckas}}\ and\ \bibinfo {author} {\bibfnamefont {Q.}~\bibnamefont
  {Nie}},\ }\bibfield  {title} {\enquote {\bibinfo {title}
  {Differentialequations. jl--a performant and feature-rich ecosystem for
  solving differential equations in julia},}\ }\href@noop {} {\bibfield
  {journal} {\bibinfo  {journal} {Journal of Open Research Software}\ }\textbf
  {\bibinfo {volume} {5}} (\bibinfo {year} {2017})}\BibitemShut {NoStop}%
\bibitem [{\citenamefont {Wanner}\ and\ \citenamefont
  {Hairer}(1996)}]{wanner1996solving}%
  \BibitemOpen
  \bibfield  {author} {\bibinfo {author} {\bibfnamefont {G.}~\bibnamefont
  {Wanner}}\ and\ \bibinfo {author} {\bibfnamefont {E.}~\bibnamefont
  {Hairer}},\ }\href@noop {} {\emph {\bibinfo {title} {Solving ordinary
  differential equations II}}}\ (\bibinfo  {publisher} {Springer Berlin
  Heidelberg},\ \bibinfo {year} {1996})\BibitemShut {NoStop}%
\end{thebibliography}%

\appendix

\section{Motivating the ILC update law}
\label{a:update}

We want to motivate the precise form of the update law used in the ILC above. If we assume that the system is roughly held in equilibrium, despite the low-amplitude
fluctuations, Eq.~\eqref{eq:mn_plant} gives us

\begin{align}
 0 &\approx - P^d_j(t) + P^{LI}_{j}(t) + P^{ILC}_{j}(t) + F_j(t) \;,
\end{align}

for each node $j = 1,...N$. In the following, we omit the node index for readability. If we neglect changes to the flows, we can approximately assume

\begin{align}
\frac{\partial P^{LI}}{\partial P^{ILC}} &\approx -1 \;,
\end{align}

i.e. a decrease in $P^{LI}$ is directly proportional to an increase in $P^{ILC}$.

The aim of the $ILC$ is adapting $P^{ILC}$ to optimize an observable $O(P^{LI})$.
Take for instance

\begin{align}
O^2(P^{LI}) &= \int\limits_0^T |P^{LI}|^2 dt \; ,
\end{align}

the quadratic norm of the lower-layer control power.
If we change $P^{ILC}$ by a constant shift $\delta P^{ILC}$ that does not depend on $t$ we 
find that the variation of $O^2(P^{LI})$ is approximated as

\begin{align}
\delta O^2(P^{LI}) &= \frac{\partial O^2(P^{LI})}{\partial P^{ILC}} \delta P^{ILC}\nonumber\\
&= 2 \left[\int\limits_0^T P^{LI} \frac{\partial P^{LI}}{\partial P^{ILC}} dt \right] \delta P^{ILC}\nonumber\\
&\approx - 2 \left[ \int\limits_0^T P^{LI} dt \right]\delta P^{ILC}\nonumber \;.
\end{align}

Thus, if we choose the $ILC$ update to be $\delta P^{ILC} = \int P^{LI} dt$ then the change in $O^2$ is always negative and we gradient descend towards a local minimum. 
As our $P^{ILC}$ are constant on the hour, this update law is sufficient to make sure that we minimize the square norm of $P^{LI}$ for each hour.

A more economic objective function could be to integrate the total cost of energy used, taking into account that there are different price points for energy bought (and scheduled) a day ahead or requested from a standing reserve of control energy.
This suggests the objective function

\begin{align}\label{eq::O_lambda}
O_\lambda (P^{LI}, P^{ILC}) &= \int\limits_0^T \left[\lambda |P^{LI}| + (1 - \lambda) |P^{ILC}| \right] dt \nonumber\\
 &= \int\limits_0^T \left[\lambda P^{LI} \sgn(P^{LI})  + (1 - \lambda) P^{ILC} \sgn(P^{ILC})  \right] dt \, ,
\end{align}

where $\lambda\in [0;1]$ is a real number. It should be noted that in reality the composition of costs is considerably more complex. For example, capacity markets reward keeping a certain amount of generation available, whether it is used or not. The objective function Eq.~\eqref{eq::O_lambda} only reflects the presence of different price levels for different levels of flexibility.

In order to calculate the variation of $O_\lambda$ with respect to a small constant shift $\delta P^{ILC}$ we  make the further assumption that the contribution from the shift in the $\sgn$ functions is of higher order, as can be expected if $P^{LI}$ is sufficiently smooth. 
Then we obtain:

\begin{align}
\delta O_\lambda &= \int\limits_0^T \lambda \sgn(P^{LI}) \delta P^{LI} dt + T (1 - \lambda) \sgn(P^{ILC}) \delta P^{ILC}\nonumber \\
&= - \left[ \lambda \int\limits_0^T \sgn(P^{LI}) dt + T (\lambda - 1) \sgn(P^{ILC})\right] \delta P^{ILC}
\end{align}

%Where $T = \int dt$. 
For $\lambda = 1$, that is day ahead energy is infinitely cheaper than instantaneous energy, we can choose the $ILC$ update

\begin{align}
\delta P^{ILC} \sim \int\limits_0^T \sgn(P^{LI}) dt
\end{align}

(where $\sim$ means "proportional to"), which guarantees that $\delta O_{\lambda = 1}$ is negative and we again descend to a sensible local minimum. 
Intuitively, this makes sense since we should increase the background power when there are more times when positive control energy is needed.

%
%In order to obtain good convergence to the local minimum we also might want to take the current scale of the discrepancy into account more explicitly as:
%
%\begin{align}
%\delta P^{ILC} \sim  \int\limits_0^T \sgn(P^{LI}) dt \,\cdot \int\limits_0^T |P^{LI}| dt
%\end{align}

For general $\lambda$, first note that as the $ILC$ compensates a positive background demand  we can always assume that $\sgn(P^{ILC}) = 1$. Then we have:

\begin{align}
\delta O_\lambda &= - \left[\lambda \int\limits_0^T \sgn(P^{LI}) dt + T (\lambda - 1)\right] \delta P^{ILC} \;.
\end{align}

Therefore, we want to chose the update law

\begin{align}
\delta P^{ILC} &\sim \left[\lambda \int\limits_0^T \sgn(P^{LI}) dt + T (\lambda - 1)\right] \;,
\end{align}

to obtain an appropriate gradient descend.

In summary we have the following economic update law:

\begin{align}
\delta P^{ILC} &= k \int\limits_0^T \left[\lambda \sgn(P^{LI}) + (\lambda - 1)\right] dt
\end{align}

% \\
%\delta P^{ILC} &= k \int\limits_0^T |P^{LI}|dt \, \cdot \int\limits_0^T \left[\lambda \sgn(P^{LI}) + (\lambda - 1)\right] dt 

\begin{figure}
\includegraphics[width=\columnwidth]{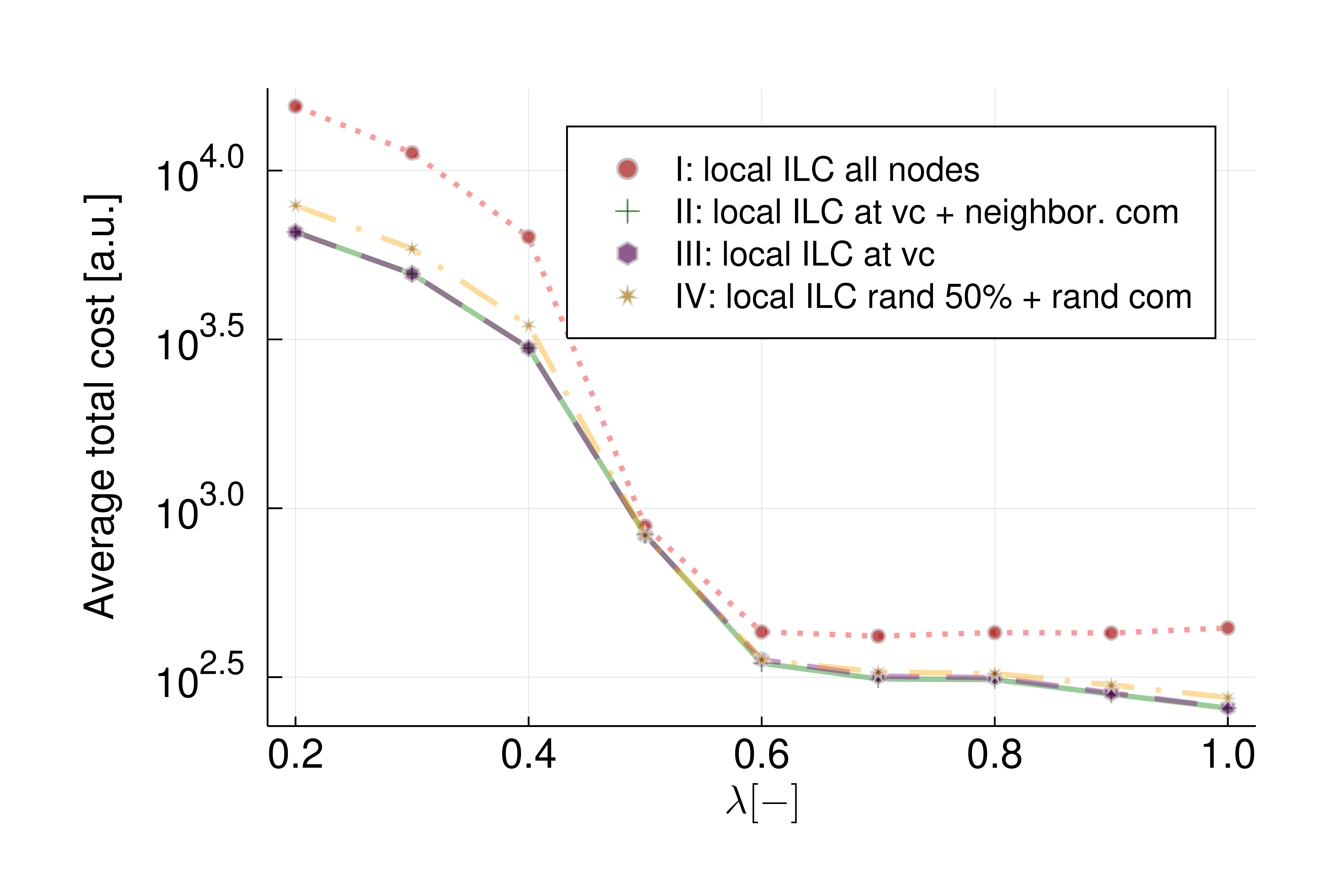}
\caption{Average mean cost over the cost factor $\lambda$ (50 days simulated, last 10 days evaluated)}
\label{fig:lambda}
\end{figure}

Fig. \ref{fig:lambda} shows the average total cost for all numerical experiments performed in this paper. $\lambda > 0.5$ are realistic scenarios, i.e. with a cheaper higher layer dispatched energy than lower layer control energy.

\section{Synthetic demand model}
\label{a:demand}

For every node $j\in\mathcal N$ in the network, 
we assume the demand is given by a periodic baseline $P^p_j$ subject to
fluctuations $P^f_j$,

\begin{eqnarray}
\label{eq:demand}
P^p_j(t) &=& A_j \sin^2\left(\pi\frac{t}{T_d}\right), \\
P^f_j(t) &=& \left(t\hspace{-2mm} \mod T_q\right) B_j\left(1 + \lfloor t / T_q\rfloor\right) + B_j\left(\lfloor t / T_q\rfloor\right),
\end{eqnarray}
where the period $T_d$ [s] is the duration of a day and the demand amplitudes $A_j \sim \mathcal{U}([0; 1])$ 
are uniform i.i.d. random numbers. The fluctuation amplitudes $B_j$ vary over time 
in a Gaussian random walk with zero mean and a variance of 0.2. $P^f_j$ is linearly interpolated 
between two consecutive updates, spaced apart by $T_q = 15$min, as given above. Note that both $A_j$ and $B_j$ are normalized with a rated power.

\section{Selection of the lower-layer control parameters}
\label{a:control_parameters}

\begin{figure}[h!]
\centering
\includegraphics[width=\columnwidth]{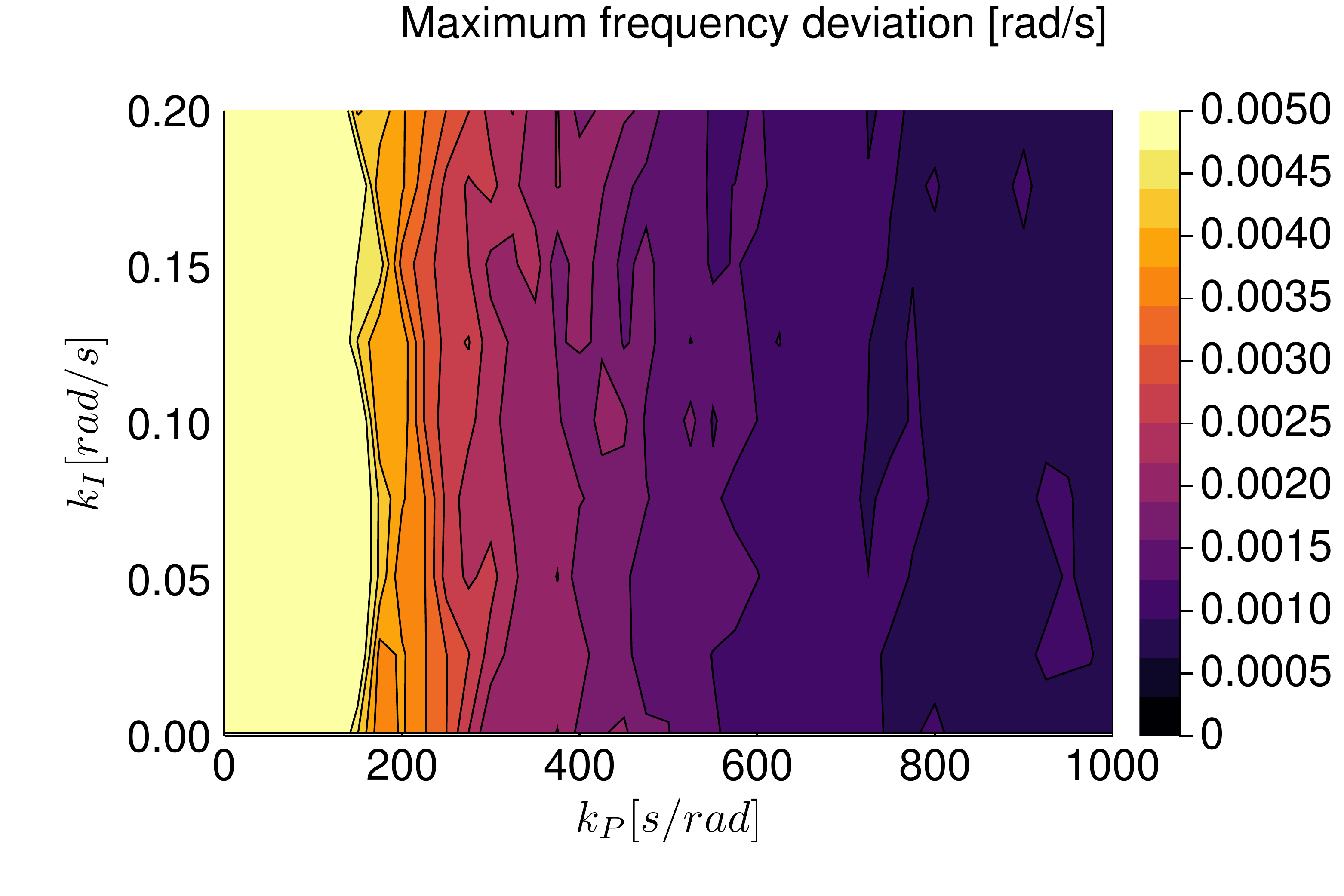}
\caption{Selection of control parameters: maximum frequency deviation; simulated and observed for 1 day, other parameters from Tab. \ref{tab:parameters} except for $k_{I,j}$ and $k_{P,j}$. The resulting maximum frequency deviation is $0.00134$ rad/s.}
\label{fig:cp_kp1}
\end{figure}

\begin{figure}[h!]
\centering
\includegraphics[width=\columnwidth]{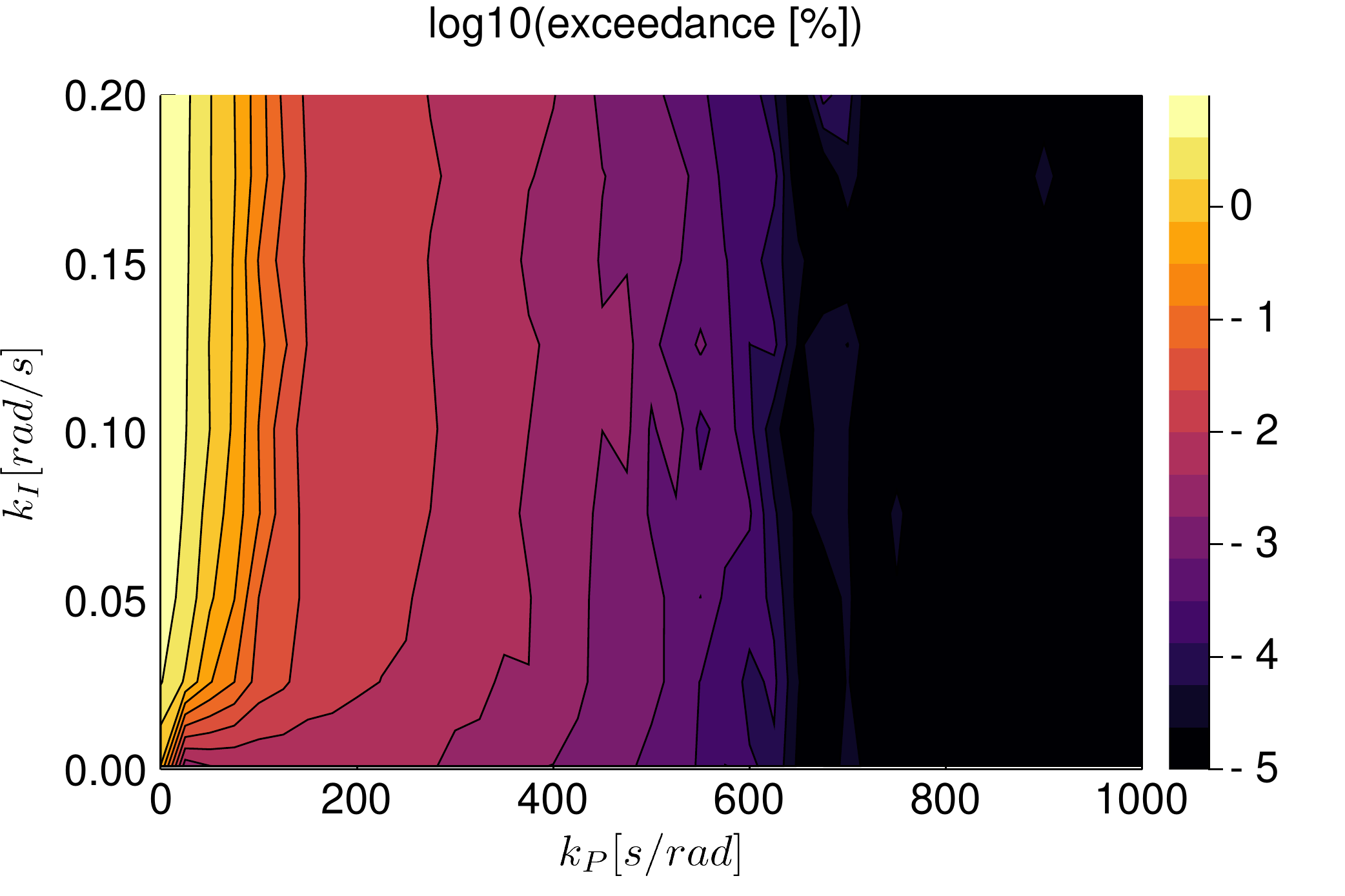}
\caption{Selection of control parameters: exceedance; simulated and observed for 1 day, other parameters from Tab. \ref{tab:parameters} except for $k_{I,j}$ and $k_{P,j}$. The resulting exceedance is 4.53 $\cdot 10^{-4}$ \%.}
\label{fig:cp_kp2}
\end{figure}

\begin{figure}[h!]
\centering
\includegraphics[width=\columnwidth]{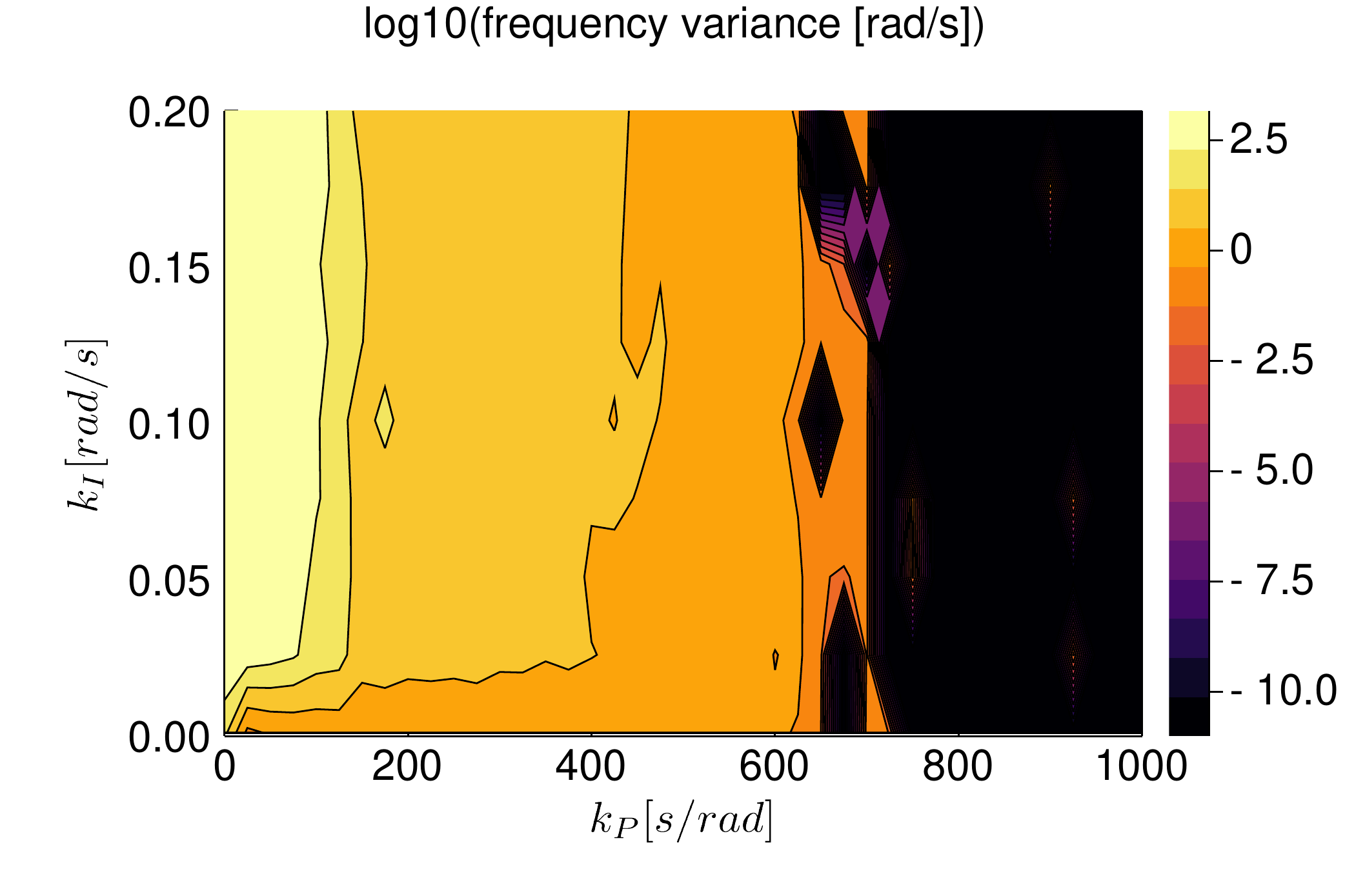}
\caption{Selection of control parameters: frequency variance; simulated and observed for 1 day, other parameters from Tab. \ref{tab:parameters} except for $k_{I,j}$ and $k_{P,j}$. The resulting frequency variance is 1.0586 rad/s.}
\label{fig:cp_kp3}
\end{figure}

The control parameters $k_{p,j}\in [0,1000]$ s/rad and $k_{I,j} \in [0.001,1]$ rad/s are varied in numerical experiments with 41 and 40 values for each parameter, respectively. This results in $41 \cdot 40 = 1640$ batches. We use 100 simulations - also called runs - in each batch. Each run has a randomly chosen 3-degree graph with 24 nodes as power network and a random demand as illustrated in Fig. \ref{fig:demand}.  Figs. \ref{fig:cp_kp1} to \ref{fig:cp_kp3} show the maximum frequency deviation, the exceedance and the frequency variance over the relevant ranges for $k_{P,j}$ and $k_{I,j}$.  We are interested in a low value for all given quantities. In combination with the analytic bounds given in the literature\cite{doerfler2018leakyintegrator}, we choose $k_{P,j} = 525$ s/rad and $k_{I,j} = 0.005$ rad/s for all nodes $j\in \mathcal N$.
\vspace{2mm}
  
\section{Representative trajectories for all scenarios}
\label{a:trajectories}

\begin{figure}[p]
\centering
\includegraphics[width=\columnwidth]{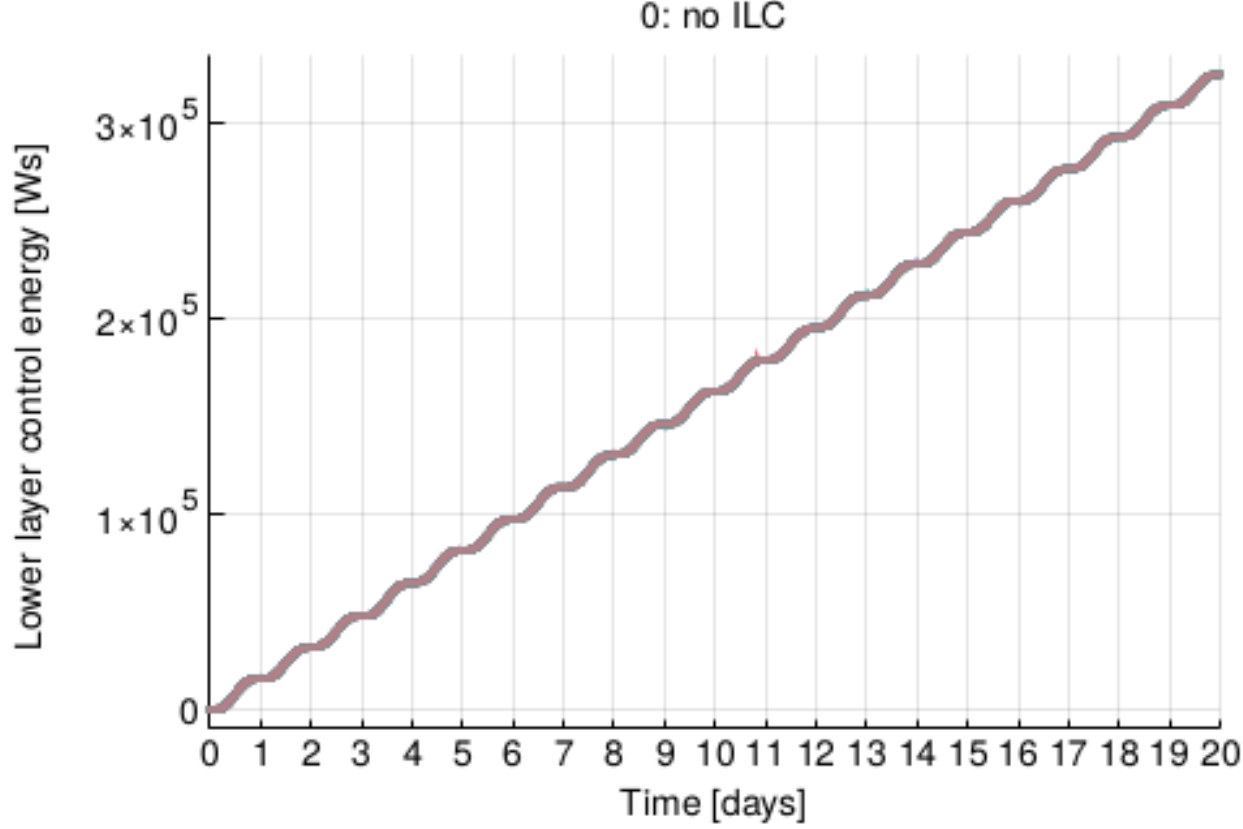}
\caption{Exemplary trajectories of lower layer control energy for 24 nodes for case 0 over a time of 20 days incl. the initial learning phase. Note that in case 0 this is the total cumulative energy over the whole time span since there are no resets.}
\label{fig:timeseries}
\end{figure}

\begin{figure}[p]
\centering
\includegraphics[width=\columnwidth]{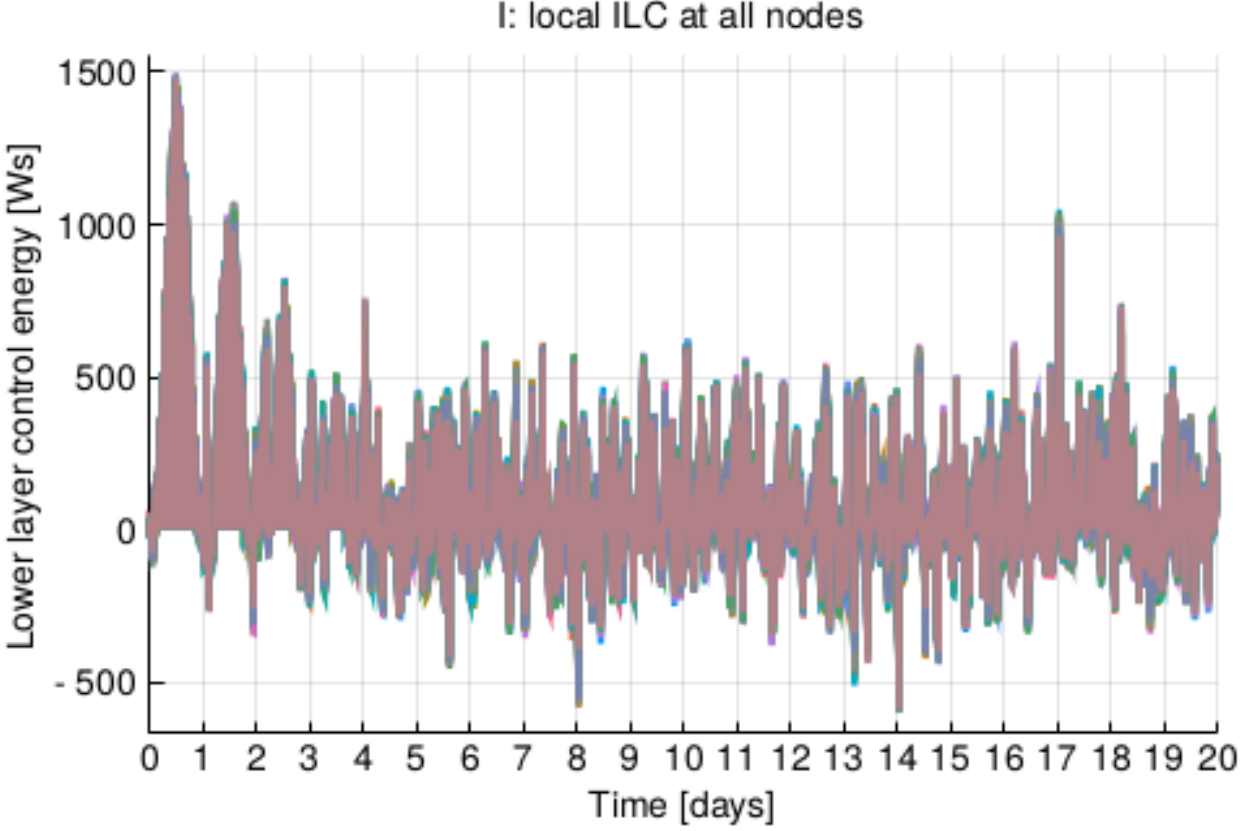}
\caption{Exemplary trajectories of lower layer control energy for 24 nodes for case I over a time of 20 days incl. the initial learning phase.}
\label{fig:timeseriesI}
\end{figure}

\begin{figure}[p]
\centering
\includegraphics[width=\columnwidth]{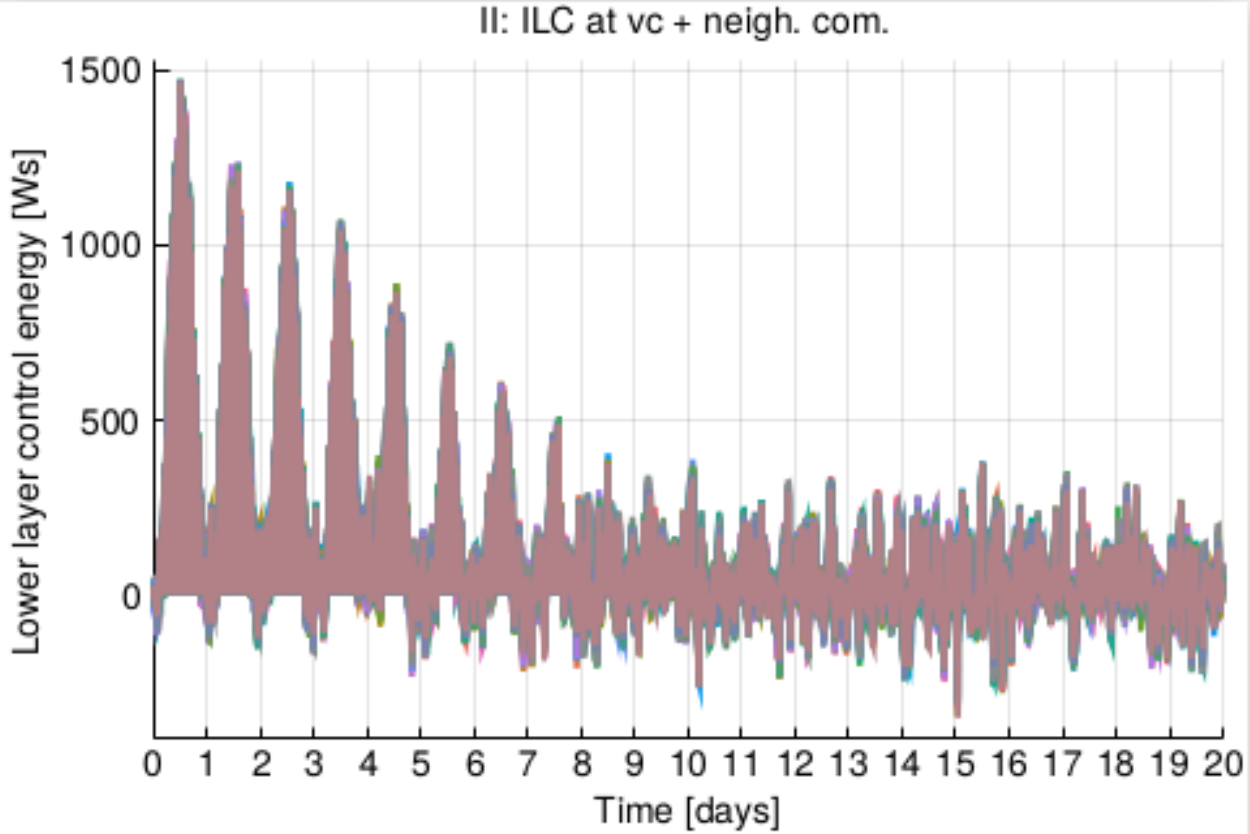}
\caption{Exemplary trajectories of lower layer control energy for 24 nodes for case II over a time of 20 days incl. the initial learning phase.}
\label{fig:timeseriesII}
\end{figure}

\begin{figure}[p]
\centering
\includegraphics[width=\columnwidth]{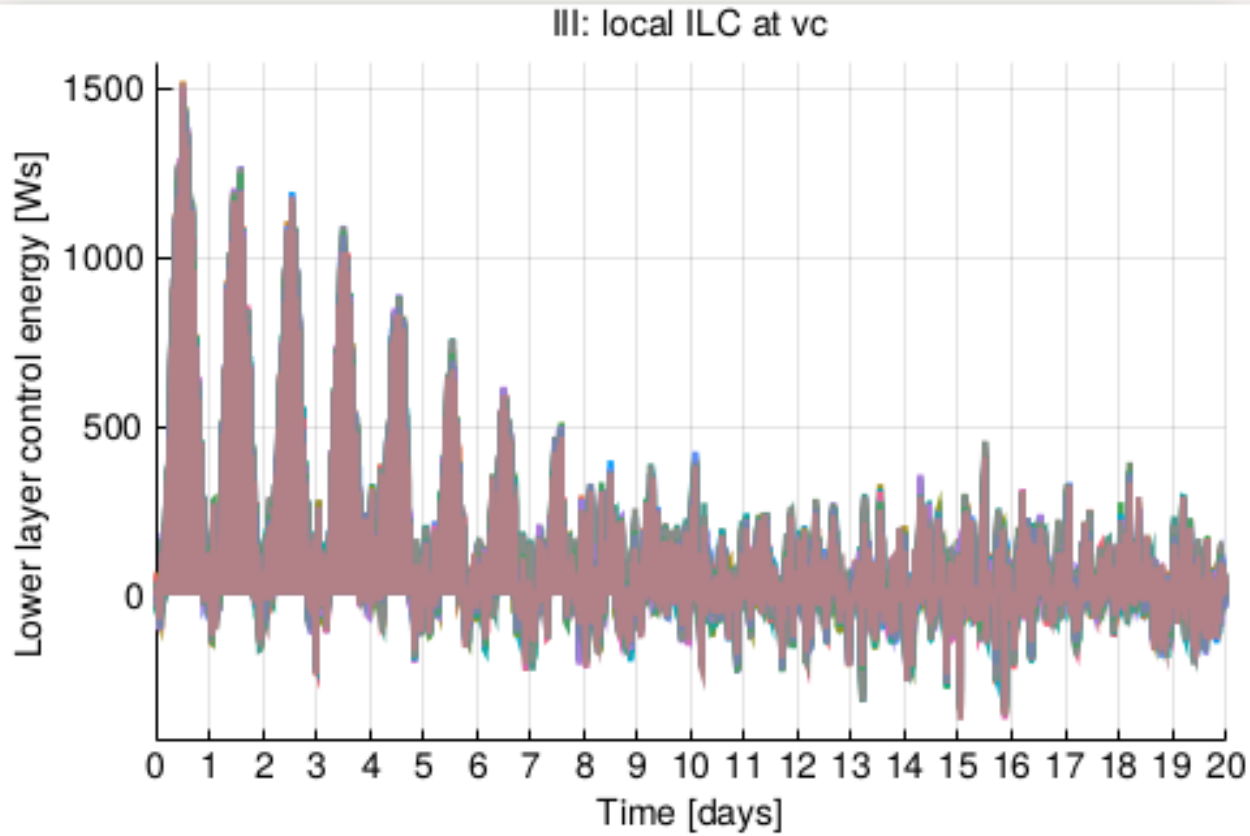}\\
\caption{Exemplary trajectories of lower layer control energy for 24 nodes for case III over a time of 20 days incl. the initial learning phase.}
\label{fig:timeseriesIII}
\end{figure}

\begin{figure}[p]
\centering
\includegraphics[width=\columnwidth]{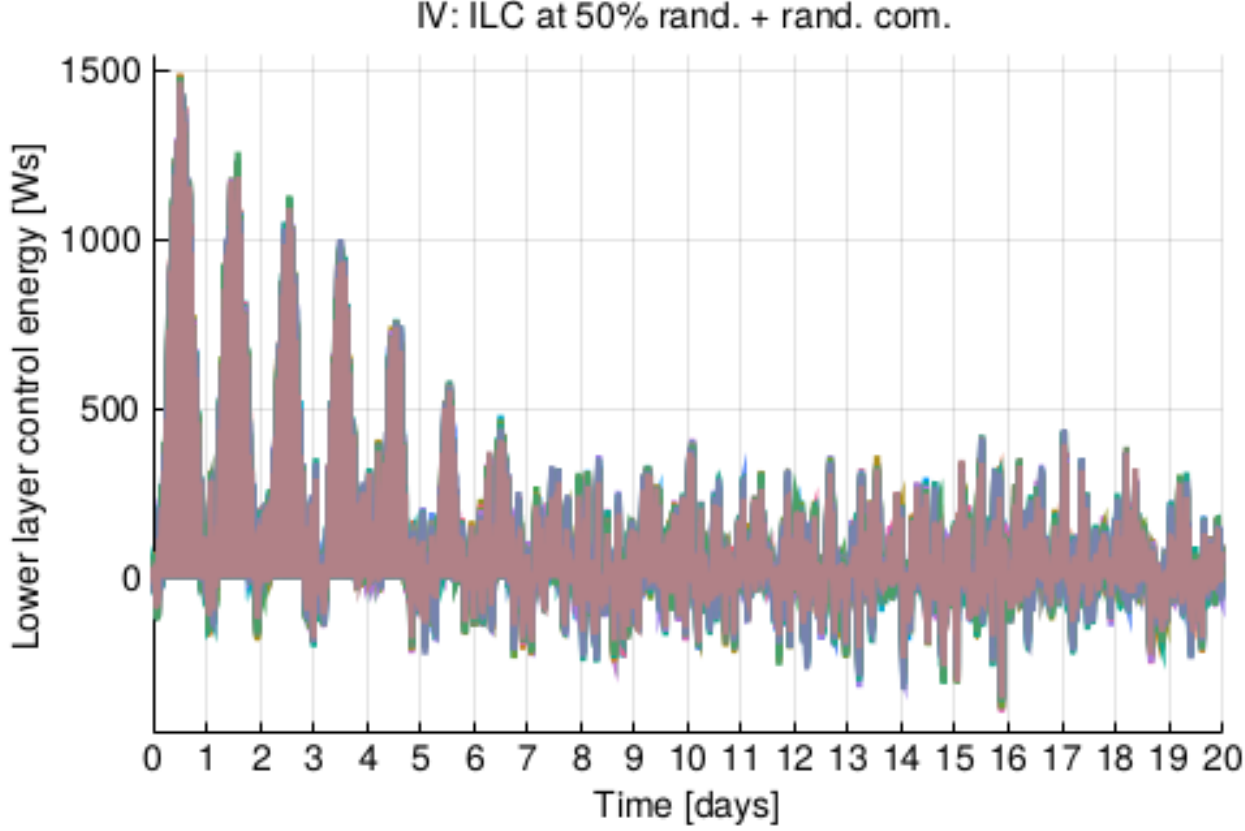}
\caption{Exemplary trajectories of lower layer control energy for 24 nodes for case IV over a time of 20 days incl. the initial learning phase.}
\label{fig:timeseriesIV}
\end{figure}

Figs. \ref{fig:timeseries} to \ref{fig:timeseriesIV}  show exemplary trajectories of cumulative lower layer control energy used for 24 nodes for cases 0-IV over a time of 20 days including the initial learning phase. In case 0 the energy is cumulative over the whole time span while in the cases I-IV it is cumulative over every hour only and then reset. Recall that for the equilibrium state analysis above, the days 20-30 are chosen which are not shown here in detail. It can be observed that ILC at all nodes learns faster but the equilibrium performance is worse than in the other cases. Other experiments performed show that during the learning process, faster learning leads to lower costs.

%\section{Variation of ILC nodes}

\end{document}